\documentclass[aps,prd,twocolumn,10pt,superscriptaddress,nofootinbib,nobibnotes,longbibliography]{revtex4-1}

\usepackage{amssymb}
\usepackage{graphicx}
\usepackage{amsmath}
\usepackage[colorlinks=true,linkcolor=blue,citecolor=blue,urlcolor=blue]{hyperref}
\usepackage{subfigure}
\usepackage{multirow}
\usepackage{setspace}
\usepackage{verbatim}
\usepackage{float}
\usepackage{color}
\usepackage{ulem}
\usepackage[utf8]{inputenc}
\usepackage[table,xcdraw]{xcolor}
\usepackage{makecell}
\usepackage{url}

\begin{document}

\title{Sound waves from primordial black hole formations}

\author{Zhuan Ning}
\email{ningzhuan17@mails.ucas.ac.cn}
\affiliation{School of Fundamental Physics and Mathematical Sciences, Hangzhou Institute for Advanced Study (HIAS), University of Chinese Academy of Sciences (UCAS), Hangzhou 310024, China}
\affiliation{Institute of Theoretical Physics, Chinese Academy of Sciences (CAS), Beijing 100190, China}
\affiliation{University of Chinese Academy of Sciences (UCAS), Beijing 100049, China}

\author{Xiang-Xi Zeng}
\email{zengxiangxi@itp.ac.cn}
\affiliation{Institute of Theoretical Physics, Chinese Academy of Sciences (CAS), Beijing 100190, China}
\affiliation{School of Physical Sciences, University of Chinese Academy of Sciences (UCAS), Beijing 100049, China}

\author{Zi-Yan Yuwen}
\email{yuwenziyan@itp.ac.cn}
\affiliation{Institute of Theoretical Physics, Chinese Academy of Sciences (CAS), Beijing 100190, China}
\affiliation{School of Physical Sciences, University of Chinese Academy of Sciences (UCAS), Beijing 100049, China}
\affiliation{Department of Physics, Stellenbosch University, Matieland 7602, South Africa}

\author{Shao-Jiang Wang}
\email{schwang@itp.ac.cn}
\affiliation{Institute of Theoretical Physics, Chinese Academy of Sciences (CAS), Beijing 100190, China}
\affiliation{Asia Pacific Center for Theoretical Physics (APCTP), Pohang 37673, Korea}

\author{Heling Deng}
\email{hd2586@columbia.edu}
\affiliation{Department of Astronomy, Columbia University, New York, NY 10027, USA}

\author{Rong-Gen Cai}
\email{caironggen@nbu.edu.cn}
\affiliation{Institute of Fundamental Physics and Quantum Technology, \& School of Physical Science and Technology, Ningbo University, Ningbo 315211, China}

\begin{abstract}
We present a numerical investigation of primordial black hole (PBH) formation from super-horizon curvature perturbations and the subsequent generation and propagation of sound waves, which can serve as a new source of stochastic gravitational wave backgrounds (SGWBs) presented in a companion letter. Using the Misner-Sharp formalism with an excision technique, our simulations extend to significantly later times than previous work and indicate that the near-critical perturbations produce a distinct compression wave featuring both overdense and underdense shells, while significantly supercritical perturbations yield only an underdense shell. We also show that a softer equation of state suppresses the formation of compression waves. Furthermore, the comoving thickness of sound shells remains nearly constant during propagation and scales with the Hubble radius at horizon re-entry, thereby serving as a key link between the gravitational-wave peak frequency and PBH mass in the companion letter. These results offer new insights into the dynamics of PBH formation and suggest potential observational signatures of PBHs in the gravitational wave (GW) spectrum from associated sound waves.
\end{abstract}

\maketitle

\section{Introduction}

In the early Universe, matter can be effectively modeled as a fluid medium capable of propagating compression waves, which manifest as sound waves in the small-amplitude regime. During cosmological first-order phase transitions (FOPTs)~\cite{Mazumdar:2018dfl,Hindmarsh:2020hop,Athron:2023xlk}, vacuum bubbles nucleate and expand, perturbing the surrounding fluid. Once they reach terminal velocity~\cite{Cai:2020djd}, the bubble walls continuously drive the fluid, generating sound waves with self-similar velocity profiles that depend solely on a dimensionless ratio between radial and temporal coordinates~\cite{Kurki-Suonio:1984zeb,Kamionkowski:1993fg,Kurki-Suonio:1995rrv,Kurki-Suonio:1995yaf}. These profiles serve as initial conditions for the sound shell model~\cite{1608.04735,1909.10040,2007.08537}, which enables semi-analytic calculations of gravitational wave (GW) spectra from the sound waves---one of the dominant sources of GWs during FOPTs~\cite{1304.2433,1504.03291}.

In addition to FOPTs, the formation of primordial black holes (PBHs)~\cite{Zeldovich:1967lct,Hawking:1971ei,Carr:1974nx} represents another violent process in the early Universe. Gravitational collapse during PBH formation induces intense interactions with ambient fluids, potentially ejecting material outwards as compression waves. In extreme cases, shock waves analogous to supernova ejecta may form, a phenomenon sometimes termed ``primordial supernovae''~\cite{1904.11482}. Collisions between these sound waves are expected to generate GWs as well. Although the sound shell model could be adapted to calculate their spectra, it requires the density and velocity profiles of the sound waves as input. In contrast to the semi-analytically derived self-similar velocity profiles from FOPTs, the configurations of PBH-induced sound waves are governed by nonlinear dynamics of gravitational collapse, necessitating full numerical simulations---the focus of this paper. The GW spectra derived from these profiles are detailed in a companion letter~\cite{Zeng:2025law}.

Beyond GW generation, PBH-induced compression waves have further cosmological implications. Carr et al.~\cite{1904.02129,1904.11482} noted that shock waves from PBH formation carry substantial energy and exhibit effective temperatures far exceeding that of the ambient plasma. These high-density hotspots could facilitate baryogenesis via electroweak sphaleron transitions~\cite{hep-ph/0310100}, naturally explaining both the observed baryon-to-photon ratio and the comparable densities of baryonic and dark matter. Although energy conservation arguments suggest that a smaller efficiency factor (thus a lighter PBH mass given the same horizon mass) may produce stronger waves, accurate numerical simulations remain essential for quantifying this energy.

Moreover, the dissipation of sound waves due to photon diffusion~\cite{Silk:1967kq} can inject energy into the background, resulting in a spectral distortion in the cosmic microwave background (CMB). 
This effect was investigated in Refs. \cite{1804.10059,2003.02485} as a result of the formation of PBHs from primordial bubbles, where the sound waves are generated by the wall-fluid interaction. It was later generalized to any PBH formation mechanisms where perturbations vanish as the spacetime approaches a flat Friedmann-Lema\^{i}tre-Robertson-Walker (FLRW) Universe asymptotically~\cite{2106.09817}.  In this case, a compensating underdense structure surrounds each PBH and propagates outwards as a sound shell. Therefore, distortions in the CMB spectrum can be used to constrain the abundance of PBHs in general. However, it remains necessary to examine the sound wave generation in other PBH formation mechanisms, such as the collapse of curvature perturbations.

PBH formation from super-horizon curvature perturbations has been extensively studied using numerical simulations, which is the primary focus of this work. In this scenario, the PBH is formed if the amplitude of horizon-scale perturbation exceeds a critical threshold. A key question in this context is whether shock waves and compression waves are generated. In Ref.~\cite{Hawke:2002rf}, the authors used a high-resolution shock-capturing (HRSC) scheme to handle potential shocks. Under sub-horizon perturbations, they detected shock waves in both slightly subcritical and supercritical evolutions, where the energy density varied by several orders of magnitude over a few grid points. Conversely, in Refs.~\cite{gr-qc/0412063,gr-qc/0605122,0811.1452}, the authors used the Hernandez-Misner formalism~\cite{Hernandez:1966zia} together with an adaptive mesh refinement (AMR) scheme and found no evidence of shock waves under super-horizon perturbations (which is more consistent with inflationary cosmology), even though the perturbation amplitudes were closer to the formation threshold than in the earlier study. They attributed this difference to the different initial conditions used. Although no supersonically propagating shock waves were detected, bounces were observed under subcritical conditions~\footnote{In this case, no black hole is formed and only the Misner-Sharp formalism~\cite{Misner:1964je} was used. The differences between the two formalisms are discussed in Section~\ref{sec:setup}.}, accompanied by compression waves propagating outwards at the speed of sound. In the supercritical case, however, the formation of compression waves was not reported. We suspect that this omission is due to the insufficient evolution time in their simulations, as compression waves emerge much later than the PBH formation. Another contributing factor may be that the physical quantities were plotted on different null slices, rendering the compression waves less apparent in the figures.

In contrast, Ref.~\cite{astro-ph/9901292} observed both bounce and compression waves in slightly supercritical cases using local time-slice plots. They found that as the perturbation amplitude increases, the bounce becomes progressively weaker and eventually disappears because the pressure gradient at the black hole horizon fails to overcome gravitational attraction. However, the above study did not quantify the shape or energy of the compression waves. In contrast to previous work that used the Hernandez-Misner formalism, in this paper, we employ the Misner-Sharp formalism combined with an excision technique to investigate PBH formation and the associated sound waves. We first reproduce the formation of compression waves along with its dynamical profile observed in the previous literature for the subcritical case, and then we demonstrate the generation of sound waves in the supercritical case, analyze their propagation and energy, and discuss their implications for GW production. 

The rest of the paper is organized as follows. In Section~\ref{sec:setup}, we describe our numerical framework. Section~\ref{sec:results} presents our simulation results and analyses of the sound waves. In Section~\ref{sec:GW}, we give an estimation of the GW energy spectrum generated by sound shell collisions. Finally, Section~\ref{sec:conclusion} contains our conclusions and discussion. Throughout this paper, units with $c = G = 1$ are used.

\section{Simulation setup} \label{sec:setup}

As with most of the other literature on PBH formation, we restrict our attention to the case of spherical symmetry (see Ref.~\cite{2111.12693} for a review). This simplification is justified by the peak theory~\cite{Bardeen:1985tr}, which suggests that large peaks tend to be almost spherically symmetric.

The Misner-Sharp~\cite{Misner:1964je} and Hernandez-Misner~\cite{Hernandez:1966zia} formalisms both describe the evolution of spherically symmetric relativistic fluids in curved spacetime using a Lagrangian formulation. The Misner-Sharp formalism employs a diagonal metric in which the time coordinate reduces to the familiar FLRW time in the absence of perturbations, commonly referred to as ``cosmic time''. This time slice is a space-like hypersurface, which is intuitive. The Misner-Sharp formalism has a well-known drawback for simulating gravitational collapse leading to black hole formation, in that singularities appear shortly after an apparent horizon forms, and the simulation breaks down, precluding the tracking of subsequent evolution (including black hole accretion and sound wave generation). However, this difficulty can be overcome by excising the region where the singularity would develop, as was done in~\cite{Escriva:2019nsa}. In contrast, the Hernandez-Misner formalism uses an outward null slicing where the time coordinate is called ``observer time''. In this coordinate system, it takes infinite coordinate time for an apparent horizon to form, thereby avoiding singularities. Consequently, previous work has mainly used this formalism to follow the PBH evolution and determine their final masses. However, we find that the Hernandez-Misner formalism develops steep gradients in the coordinate space during the later stages of numerical simulations~\cite{Bloomfield:2015ila}. Resolving these features requires a computationally expensive AMR scheme. Therefore, we adopt the Misner-Sharp formalism with an excision technique for its intuitive cosmic-time framework and capacity to probe post-PBH sound wave generation.

\subsection{Misner-Sharp formalism}

In the Misner-Sharp formalism, the line element for a spherically symmetric spacetime is given by
\begin{equation}
  ds^2=-A(t,r)^2\mathrm{d}t^2+B(t,r)^2\mathrm{d}r^2+R(t,r)^2\mathrm{d}\Omega^2,
\end{equation}
where $\mathrm{d}\Omega^2=\mathrm{d}\theta^2+\sin^2\theta \mathrm{d}\phi^2$ represents the line element of the unit sphere, and $R(t,r)$ is the areal radius. We consider a perfect fluid described by the energy-momentum tensor
\begin{equation} \label{eq:energy-momentum}
  T_{\mu\nu}=(\rho+p)u_\mu u_\nu+pg_{\mu\nu},
\end{equation}
where $\rho$ denotes the energy density, $p$ the pressure, and $u^\mu=A^{-1}\delta^\mu_0$ the comoving four-velocity of the fluid. Following the notation in Ref.~\cite{Misner:1964je}, we introduce the following auxiliary variables:
\begin{subequations}
  \begin{align}
    D_tR &\equiv \frac{1}{A}\frac{\partial R}{\partial t} \equiv U(t,r),\\
    D_rR &\equiv \frac{1}{B}\frac{\partial R}{\partial r} \equiv \Gamma(t,r),
  \end{align}
\end{subequations}
where $D_t$ and $D_r$ are the proper time and distance derivatives, respectively. Here, $U$ represents the radial component of the four-velocity in the associated Eulerian frame, measuring the radial velocity of the fluid relative to the coordinate center, and $\Gamma$ is a generalization of the Lorentz factor. The Misner-Sharp mass is defined as
\begin{equation}
  M(t,r) \equiv \int_0^R4\pi R^2\rho\left(\frac{\partial R}{\partial r}\right)\mathrm{d}r,
\end{equation}
which quantifies the mass-energy contained within a sphere of coordinate radius $r$.

Using these variables, the Einstein equations and energy-momentum conservation ($\nabla^\mu T_{\mu\nu} = 0$) yield:
\begin{subequations}
  \begin{align}
    \label{eq:A} D_rA&=\frac{-A}{\rho+p}D_rp,\\
    \label{eq:Hamiltonian} D_rM&=4\pi\Gamma\rho R^2,\\
    \label{eq:Gamma} \Gamma&=\sqrt{1+U^2-\frac{2M}{R}},\\
    D_tR&=U,\\
    D_tU&=-\left[\frac{\Gamma}{(\rho+p)}D_rp+\frac{M}{R^2}+4\pi Rp\right],\\
    D_tM&=-4\pi R^2Up,\\
    D_t\rho&=-\frac{(\rho+p)}{\Gamma R^2}D_r(UR^2).
  \end{align}
\end{subequations}
Eq.~\eqref{eq:Hamiltonian} is the Hamiltonian constraint, which we use to monitor numerical errors (see Appendix~\ref{app:convergence}). The remaining equations govern the temporal evolution of the gravitational system. The boundary conditions are $R(t,r=0) = 0$, $U(t,r=0) = 0$, $M(t,r=0) = 0$, and $D_r\rho(t,r=0) = 0$ for spherical symmetry, along with an arbitrary nonzero boundary condition for $A$.

Using Eq.~\eqref{eq:Gamma}, the Misner-sharp mass can alternatively be expressed as
\begin{equation}
  M(t,r)=\int_0^r\rho\sqrt{1+U^2-\frac{2M}{R}}\mathrm{d}V,
\end{equation}
where $\mathrm{d}V \equiv 4\pi R^2B\mathrm{d}r$ is the proper volume element. Therefore, $M$ includes contributions from both kinetic and gravitational potential energy. We use $M$ to quantify the energy of sound waves in Section~\ref{sec:energy}.

\subsection{Cosmological setup}

In the cosmological context, we consider the equation of state $p=\omega\rho$ with constant $\omega$ to close the system of equations. By requiring spatial flatness and a homogeneous background energy density (i.e., $\partial_r \rho_b = 0$, where the subscript $b$ denotes the background FLRW value), we can get the solution for the background spacetime:
\begin{subequations}
  \begin{align}
    A_b &= 1, \\
    \Gamma_b &= 1, \\
    \rho_b(t)&=\rho_0 a(t)^{-3(1+\omega)}, \\
    R_b(t,r) &= a(t)r, \\
    U_b(t,r) &= H(t)R_b(t,r), \\
    M_b(t,r) &= \frac{4\pi}{3}\rho_b(t) R_b(t,r)^3,
  \end{align}
\end{subequations}
where $\rho_0$ is the initial energy density, $a(t)$ is the scale factor, and $H(t)=\dot{a}/a$ is the Hubble parameter. Setting the initial time to $t_0$ with $a(t_0)=1$ yields
\begin{subequations}
  \begin{align}
    \rho_0 &= \frac{3H(t_0)^2}{8\pi} = \frac{3\alpha^2}{8\pi t_0^2}, \\
    a(t) &= \left(\frac{t}{t_0}\right)^\alpha, \\
    H(t) &= \frac{\alpha}{t},
  \end{align}
\end{subequations}
with $\alpha = 2/3(1 + \omega)$ ($\alpha = 1/2$ for a radiation-dominated era where $\omega = 1/3$).

For solutions beyond the background, we impose $A(t,r_f)=1$ to match the FLRW spacetime outside the computational domain, where $r_f$ is the outermost grid point. Then Eq.~\eqref{eq:A} can be integrated to obtain
\begin{equation}
  A(t,r) = \left(\frac{\rho_b(t)}{\rho(t,r)}\right)^{\omega/(1+\omega)}
\end{equation}

Conventional numerical approaches typically evolve the primitive variables $R$, $U$, $M$, and $\rho$ directly. In this work, we need to simulate the system over a much longer time to study the formation and propagation of sound waves. Therefore, to enhance numerical accuracy and stability, we follow Refs.~\cite{Polnarev:2012bi,Bloomfield:2015ila} to factor out the background evolution from the dynamical variables. Specifically, we redefine
\begin{subequations} \label{eq:redefined}
  \begin{align}
    \rho &= \rho_b\bar{\rho}, \\
    p &= \rho_b\bar{p}, \\
    R &= R_b\bar{R} = ar\bar{R}, \\
    U &= HR\bar{U}, \\
    M &= \frac{4\pi}{3}\rho_bR^3\bar{M},
  \end{align}
\end{subequations}
where the barred variables $\bar{X}$ (with $X = \rho,R,M,U$) equal unity for the FLRW solution and are even functions of $r$. We scale $M$ and $U$ by $R$ instead of $R_b$ because the expected quantities depend on the physical radius rather than the unperturbed FLRW radius.

To accelerate the numerical evolution, we introduce the logarithmic time variable $\xi \equiv \ln(t/t_0)$, so that $\partial_t = \frac{H}{\alpha}\partial_\xi$. Additionally, we define the dimensionless coordinate radius $\bar{r} \equiv r/R_H$, where $R_H = H^{-1}(t_0) = t_0/\alpha$ is the initial Hubble radius. In this case, all quantities are dimensionless and our plots are independent of the choice of $R_H$. With these new variables, the equations of motion transform to:
\begin{subequations} \label{eq:eom}
  \begin{align}
    A &= \bar{\rho}^{-3\alpha\omega/2},\\
    \bar{\rho} &= \bar{M}+\frac{\bar{r}\bar{R}}{3(\bar{r}\bar{R})'}\bar{M}',\\
    \bar{\Gamma}^2 &\equiv \frac{\Gamma^2}{a^2H^2R_H^2} = e^{2(1-\alpha)\xi}+\bar{r}^2\bar{R}^2\left(\bar{U}^2-\bar{M}\right),\\
    \partial_{\xi}\bar{R} &= \alpha\bar{R}\left(\bar{U}A-1\right),\\
    \partial_{\xi}\bar{U} &= \bar{U}-\alpha A\bigg[\bar{\Gamma}^2\frac{\bar{p}^{\prime}}{\bar{r}\bar{R}(\bar{R}+\bar{r}\bar{R}^{\prime})(\bar{\rho}+\bar{p})}\nonumber\\
    &\quad+\frac12\left(2\bar{U}^2+\bar{M}+3\bar{p}\right)\bigg],\\
    \partial_\xi\bar{M} &= 2\bar{M}-3\alpha\bar{U}A\left(\bar{p}+\bar{M}\right),\\
    \partial_\xi\bar{\rho} &= -\alpha(\bar{\rho}+\bar{p})A\left(3\bar{U}+\frac{\bar{r}\bar{R}\bar{U}'}{\bar{R}+\bar{r}\bar{R}'}\right) + 3\alpha(1+\omega)\bar{\rho},
  \end{align}
\end{subequations}
where the primes denote derivatives with respect to $\bar{r}$. The boundary conditions require $\bar{X}' = 0$ at the origin $\bar{r} = 0$, implying that all quantities are even functions of $\bar{r}$. At the outer boundary $\bar{r} = \bar{r}_f$, all dynamical variables match the FLRW solution, i.e., $\bar{X} = 1$.

\subsection{Initial conditions}

For perturbations at super-horizon scales, we use the long-wavelength approximation~\cite{Salopek:1990jq} (which is also called the gradient expansion method) to set up consistent initial conditions. We consider a cosmological perturbation with an initial length scale $r_i$ much larger than the Hubble horizon radius $H^{-1}$ and introduce a parameter $\epsilon$ to relate these two scales:
\begin{equation}
  \epsilon(t) \equiv \frac{H^{-1}(t)}{a(t)r_i} = \frac{1}{a(t)H(t)r_i} \ll 1.
\end{equation}
In this approximation, inhomogeneities are expanded in terms of $\epsilon(t)$. At the leading order, the perturbed spacetime can be expressed as
\begin{equation}
  \mathrm{d}s^2 = -\mathrm{d}t^2 + a^2(t)\left(\frac{\mathrm{d}r^2}{1-K(r)r^2}+r^2\mathrm{d}\Omega^2\right),
\end{equation}
where the curvature perturbation $K(r)$ encodes the deviation from homogeneity. Following Ref.~\cite{gr-qc/0605122}, the Misner-Sharp equations are solved perturbatively:
\begin{subequations}
  \begin{align}
  R(t,r)&=a(t)r\left(1+\epsilon^2(t)\tilde{R}(r)\right),\\
  U(t,r)&=H(t)R(t,r)\left(1+\epsilon^2(t)\tilde{U}(r)\right),\\
  M(t,r)&=\frac{4\pi}3\rho_b(t)R(t,r)^3\left(1+\epsilon^2(t)\tilde{M}(r)\right),\\
  \rho(t,r)&=\rho_b(t)\left(1+\epsilon^2(t)\tilde{\rho}(r)\right).
  \end{align}
\end{subequations}
In the limit $\epsilon \rightarrow 0$, the FLRW solution is recovered. The tilde variables have been computed in Refs.~\cite{gr-qc/0605122,1809.02127} and are given by
\begin{subequations}
  \begin{align}
    \tilde{U}(r)&=-\frac1{5+3\omega}K(r)r_i^2,\\
    \tilde{M}(r)&=-3(1+\omega)\tilde{U}(r),\\
    \tilde{\rho}(r)&=\frac{3(1+\omega)}{5+3\omega}\left[K(r)+\frac r3K'(r)\right]r_i^2,\\
    \tilde{R}(r)&=-\frac \omega{(1+3\omega)(1+\omega)}\tilde{\rho}(r)+\frac1{1+3\omega}\tilde{U}(r).
  \end{align}
\end{subequations}
Accordingly, the initial conditions for the redefined variables $\bar{X}$ are set by
\begin{align}
  \bar{X}(t_0,r)=1+\epsilon^2(t_0)\tilde{X}(r).
\end{align}

The initial perturbation length scale $r_i$ is identified as the location where the compactness function~\cite{gr-qc/9905064} (which quantifies the mass excess within a given volume),
\begin{equation}
  \mathcal{C}(t,r) \equiv \frac{2\left[M(t,r)-M_b(t,r)\right]}{R(t,r)},
\end{equation}
attains its maximum. At the leading order of the gradient expansion, we have
\begin{equation}
  \mathcal{C}(t,r)\simeq \mathcal{C}(r)=f(\omega)K(r)r^2,
\end{equation}
with $f(\omega) = 3(1+\omega)/(5+3\omega)$. Then $r_i$ is related to the curvature perturbation $K(r)$ by $\mathcal{C}'(r_i)=0$, which gives
\begin{equation}
  K(r_i)+\frac{r_i}{2}K^{\prime}(r_i)=0.
\end{equation}

The time scale $t_m$, defined by $\epsilon(t_m) = 1$, corresponds to the time when the physical length scale of the perturbation equals the Hubble horizon radius---that is, the horizon crossing time:
\begin{equation}
  t_m = (r_i/R_H)^{1/(1-\alpha)}t_0.
\end{equation}
The horizon mass $M_H$ at $t_m$ is used as the unit for black hole mass and is given by $M_H = 1/2H(t_m) = t_m/2\alpha$.

To quantify the amplitude of a cosmological perturbation, the averaged mass excess within a certain volume $V = 4\pi R^3/3$ is defined as
\begin{equation}
  \bar{\delta}(t,r) \equiv \frac{1}{V}\int_0^r 4\pi R^2\frac{\rho-\rho_b}{\rho_b}\mathrm{d}R,
\end{equation}
At the leading order in terms of $\epsilon$, this expression becomes
\begin{equation}
  \bar{\delta}(t,r) = \epsilon^2(t)\tilde{\delta}(r),
\end{equation}
with $\tilde{\delta}(r) = f(\omega)K(r)r_i^2 = r_i^2\mathcal{C}(r)/r^2$. The perturbation amplitude is then defined as
\begin{equation}
  \delta_m \equiv\bar{\delta}(r_i,t_m) = \tilde{\delta}(r_i) = f(\omega)K(r_i)r_i^2 = \mathcal{C}(r_i).
\end{equation}

\subsection{Apparent horizon formation} \label{sec:ah}

We use the appearance of an apparent horizon as the criterion for black hole formation. An apparent horizon is defined as a marginally trapped surface, which requires that the expansion of one of the two null geodesic congruences orthogonal to this surface vanishes while the other is negative. Let $k^\pm$ denote the outgoing ($+$) and ingoing ($-$) null vectors, given by
\begin{equation}
  k_\mu^\pm = \frac{1}{\sqrt{2}}(-A,\pm B,0,0),
\end{equation}
so that $k^+ \cdot k^- = -1$. The expansion $\Theta^\pm$ is defined as
\begin{equation}
  \Theta^\pm \equiv h^{\mu\nu}\nabla_\mu k_\nu^\pm,
\end{equation}
where $h^{\mu\nu}$ is the induced metric on the spherical surface. In the Misner-Sharp formalism, this expression simplifies to
\begin{equation}
  \Theta^\pm = \frac{2}{R}(U\pm\Gamma).
\end{equation}
A marginally trapped surface forms when $\Theta^+ = 0$ and $\Theta^- < 0$. Since $\Gamma$ is always positive, the condition $\Theta^+ = 0$ implies that $U$ must be negative, which ensures that $\Theta^- < 0$ is satisfied. Using Eq.~\eqref{eq:Gamma}, the condition for apparent horizon formation becomes
\begin{equation}
  2M = R.
\end{equation}
In terms of the redefined variables, this condition is expressed as
\begin{equation}
  \bar{R}^2\bar{M}\bar{r}^2e^{2(\alpha-1)\xi} = 1.
\end{equation}
Thus, an apparent horizon is detected when this relation holds and $\bar{U} < 0$.

Once an apparent horizon forms, the region inside it can be safely excised, as the evolution of matter within the horizon does not affect the exterior. Since the apparent horizon moves outwards over time, its motion is tracked, and the grid is truncated repeatedly. This procedure, known as the excision technique, is detailed in the following subsection.

\subsection{Numerical implementation}

To solve the Misner-Sharp equations numerically, we discretize the radial coordinate $\bar{r}$ using a fourth-order central difference scheme. Before black hole formation, the computational domain extends from the origin to an outer boundary $r_f$, chosen to be much larger than the initial perturbation scale $r_i$. A staggered grid is used to avoid the coordinate singularity at the origin. Specifically, we define a uniform grid as
\begin{equation}
  \bar{r}_i = \left(i-\frac{1}{2}\right)\Delta \bar{r},\quad i = 1,2,\dots,N,
\end{equation}
where $N$ is the number of grid points and the grid spacing is $\Delta \bar{r} = r_f/R_H(N-1/2)$. As in the publicly available code \texttt{SPriBoSH}~\cite{Escriva:2019nsa}, all dynamical variables are defined at the grid points. To impose boundary conditions, we introduce two ghost cells at the inner and outer boundaries (denoted as $\bar{r}_{-2}$, $\bar{r}_{-1}$, $\bar{r}_{N+1}$, and $\bar{r}_{N+2}$). At the inner boundary, we require the redefined variables to be even functions of $\bar{r}$, i.e., $\bar{X}(\bar{r}_{-2}) = \bar{X}(\bar{r}_2)$ and $\bar{X}(\bar{r}_{-1}) = \bar{X}(\bar{r}_1)$. At the outer boundary, the solution is matched to the FLRW background, so $\bar{X}(r_{N+1}) = \bar{X}(r_{N+2}) = 1$.

In the time direction, we use the fourth-order Runge-Kutta method to update the variables. The time step is constrained by the Courant-Friedrichs-Lewy (CFL) condition:
\begin{equation} \label{eq:CFL}
  c_s=\sqrt{\omega}<\frac{B\Delta r}{A\Delta t},
\end{equation}
where $c_s$ is the speed of sound, $\Delta r$ is the radial coordinate spacing, $\Delta t$ is the time step, and the metric component $B$ is given by
\begin{equation}
  B = \frac{\partial_rR}{\Gamma} = \frac{\bar{R}+\bar{r}\bar{R}}{\bar{\Gamma}}e^\xi.
\end{equation}
This condition can be rewritten as
\begin{equation}
  \Delta t<\frac{t_0}{\alpha\sqrt{\omega}}\frac{B\Delta \bar{r}}{A}.
\end{equation}
Expressed in logarithmic time $\xi$, the time step satisfies
\begin{equation}
  \Delta\xi=\ln\left(1+\frac{\Delta t}t\right)<\ln\left(1+\frac{e^{-\xi}}{\alpha\sqrt{\omega}}\frac{B\Delta\bar{r}}{A}\right).
\end{equation}

At each time step, the formation of an apparent horizon is monitored according to the criteria in Section~\ref{sec:ah}. Once an apparent horizon is detected, its location is determined via cubic spline interpolation, and an excision procedure is applied to remove the interior region, thereby preventing numerical breakdown due to singularity formation. This strategy, similar to that in Ref.~\cite{Escriva:2019nsa}, involves reconstructing the grid with the apparent horizon as the new inner boundary and interpolating the dynamical variables onto the revised grid. The evolution then continues until the next excision.

Following excision, inner ghost cells are omitted, and a fourth-order one-sided difference scheme is employed. Although there is no physical boundary condition on the excision surface, we find that freezing the derivative of $\bar{\rho}$ there, after each grid redefinition, enhances numerical stability without affecting the overall results.

After apparent horizon formation, the gradients of the dynamical variables near the horizon become very steep. To resolve these gradients, we use a logarithmic grid to enhance resolution near the horizon. Specifically, the new grid is defined as
\begin{equation}
  \bar{r}_i = \bar{r}_{\mathrm{AH}} + \Delta \bar{r}_0\frac{C^{i-1}-1}{C-1},\quad i=1,2,\dots,N,
\end{equation}
where $\bar{r}_{\mathrm{AH}}$ is the apparent horizon location, $\Delta \bar{r}_0$ is the first grid spacing, and $C$ is a constant slightly larger than $1$. We find $C=1+9/N$ to be sufficient for resolving the formation and propagation of sound waves.

With these numerical methods, no artificial viscosity is required to stabilize the code. The implementation is carried out in Python using the \texttt{Numpy} and \texttt{Scipy} libraries for numerical computations.

\section{Numerical results} \label{sec:results}

\subsection{Formation of compression waves}

In this subsection, we investigate the factors that influence the generation of compression waves during PBH formation. We begin by focusing on the radiation-dominated era ($\omega = 1/3$) and examining the effect of varying perturbation amplitudes. To this end, we consider a Gaussian curvature perturbation of the form
\begin{equation} \label{eq:perturbation}
  K(r) = \mathcal{A}\bar{K}(r) = \mathcal{A} e^{-(r/r_i)^2},
\end{equation}
where $r_i$ is the initial perturbation scale and the peak amplitude $\mathcal{A}$ is related to the perturbation amplitude $\delta_m$ by
\begin{equation}
    \mathcal{A} = \frac{\delta_m}{f(\omega)\bar{K}(r_i)r_i^2} = \frac{(5+3\omega)\delta_m}{3(1+\omega)r_i^2}.
\end{equation}
As demonstrated in Ref.~\cite{1809.02127}, this curvature perturbation results in a compensated energy density perturbation in which the mass excess of the central overdensity is exactly balanced by a surrounding underdensity that asymptotically approaches the background density.

Without loss of generality, we set $r_i = 10R_H$ to satisfy the long-wavelength approximation. Consequently, the time of horizon crossing is $t_m = 100t_0$, and the horizon mass at $t_m$ is $M_H = 100t_0$. The outer boundary is chosen as either $r_f = 150R_H$ or $r_f = 300R_H$, depending on the case. We employ $4000$ grid points and use a logarithmic grid with the parameter $C = 1 + 9/N$. By applying a dichotomy method, we determine the critical perturbation amplitude to be $\delta_c \approx 0.49774\pm1\times10^{-5}$, in agreement with the results of Ref.~\cite{Escriva:2019nsa}. This outcome validates our code, and a convergence test is presented in Appendix~\ref{app:convergence}.

\begin{figure*}[htbp]
  \centering
  \includegraphics[width=0.45\linewidth]{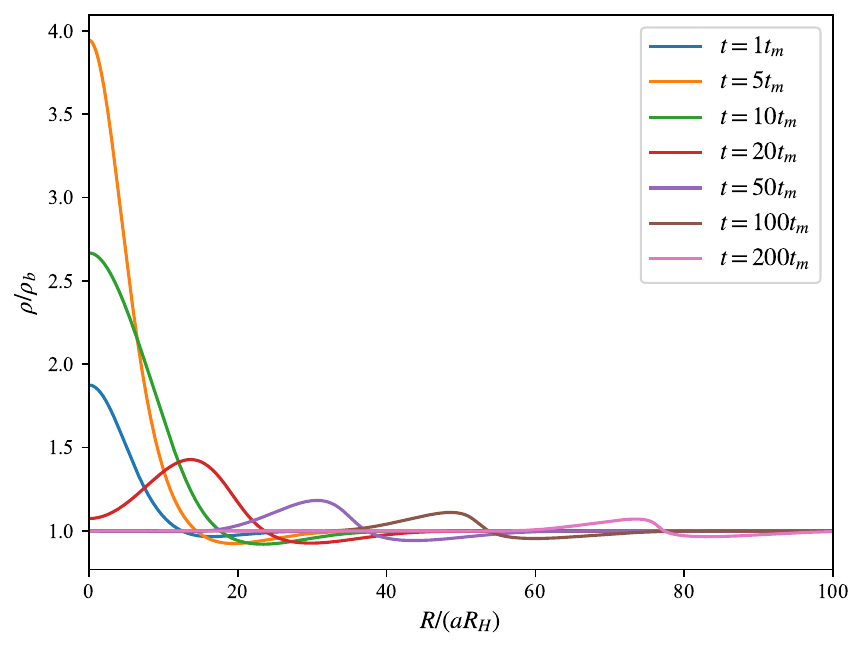}
  \qquad
  \includegraphics[width=0.45\linewidth]{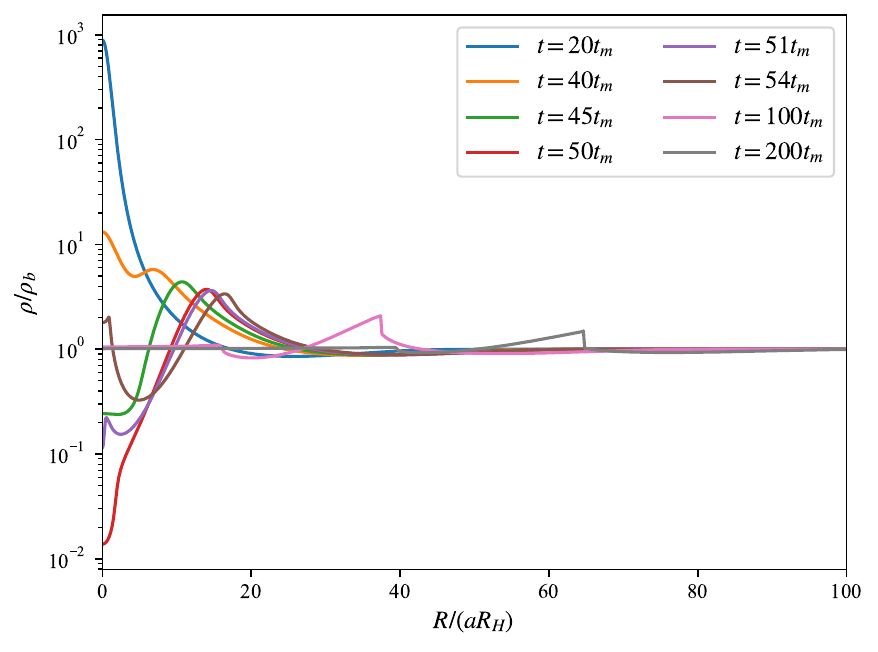}
  \caption{Time evolution of the energy density for subcritical perturbations with $\delta_m = 0.3 \ll \delta_c$ (left panel) and $\delta_m = 0.495 \lesssim \delta_c$ (right panel). The horizontal coordinate $R/a$ indicates the comoving radius, and different coloured lines correspond to different time snapshots.}
  \label{fig:subcritical}
\end{figure*}

For subcritical perturbations, previous studies~\cite{gr-qc/0412063,gr-qc/0605122} have thoroughly investigated the hydrodynamical evolution. They found that the perturbation initially grows and then bounces back to form a compression wave propagating outwards. When $\delta_m$ is very close to $\delta_c$, a second, more violent bounce is observed. Our simulation results for $\delta_m = 0.3 \ll \delta_c$ and $\delta_m = 0.495 \lesssim \delta_c$ (see Fig.~\ref{fig:subcritical}) are consistent with these findings. For $\delta_m = 0.495$, the first bounce causes the central density to drop below the background level (as indicated by the green and red lines); the resulting rarefaction region is rapidly refilled (purple line), leading to a second bounce (brown line). In contrast, for $\delta_m = 0.3$, the central density remains above the background value, and only one bounce occurs.

\begin{figure*}[htbp]
  \centering
  \includegraphics[width=0.45\linewidth]{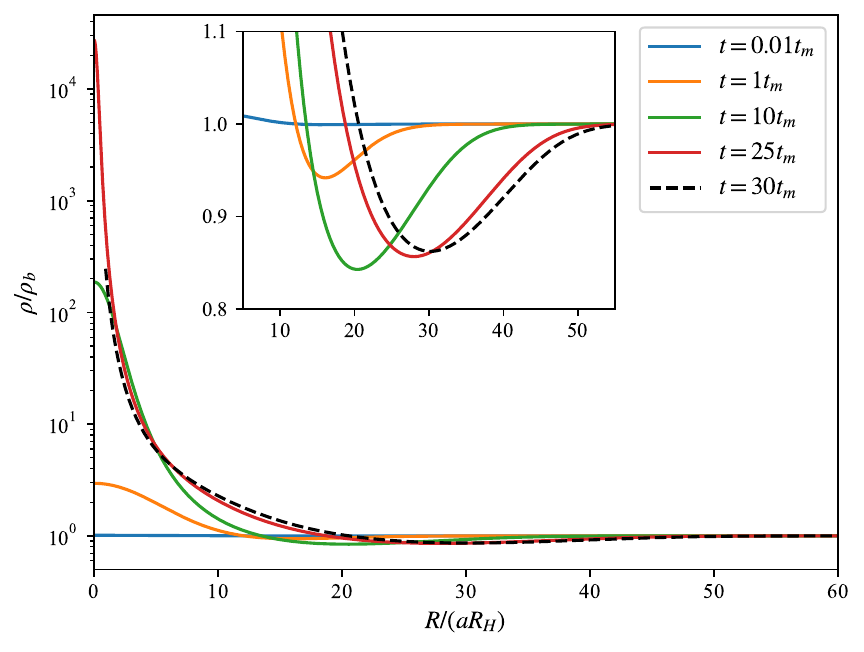}
  \qquad
  \includegraphics[width=0.45\linewidth]{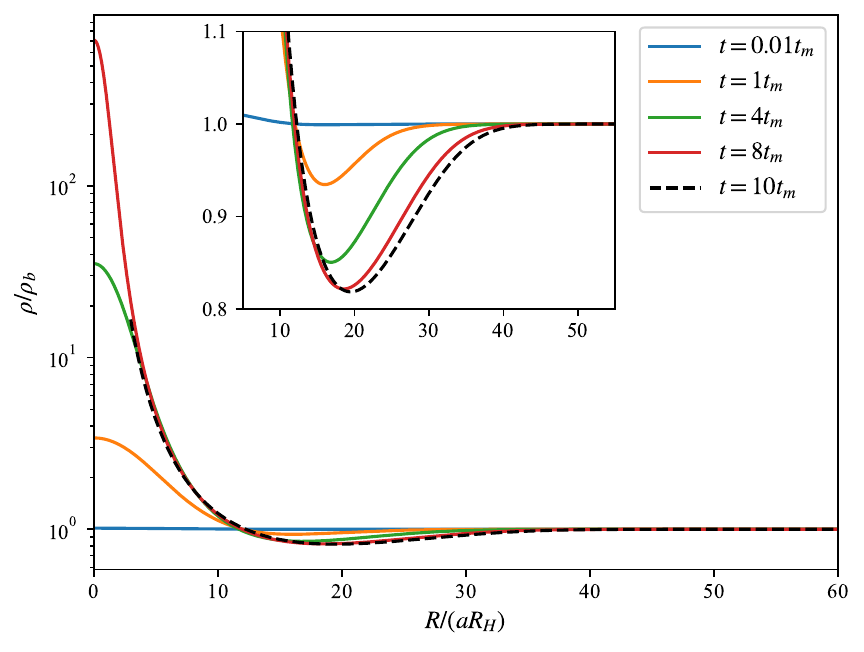}
  \caption{Time evolution of the energy density before PBH formation for near-critical ($\delta_m = 0.5$, left panel) and far-from-critical ($\delta_m = 0.55$, right panel) perturbations. The black dashed lines indicate that a black hole has formed.}
  \label{fig:before}
\end{figure*}

We focus on supercritical perturbations in the following, which have not been investigated in previous work. Two representative cases are considered: a near-critical perturbation ($\delta_m = 0.5$) and a far-from-critical perturbation ($\delta_m = 0.55$). Before PBH formation, the time evolution of the energy density is qualitatively similar for both cases, as shown in Fig.~\ref{fig:before}. In each case, the central overdensity increases rapidly while the outer underdensity gradually widens to compensate for it. The black dashed lines in the diagrams indicate that a black hole has formed, as its interior is excised. Notably, for near-critical perturbations, PBH formation occurs later and the central overdensity reaches larger amplitudes than the far-from-critical case, implying a stronger pressure gradient. This leads to a different subsequent evolution.

\begin{figure*}[htbp]
  \centering
  \includegraphics[width=0.45\linewidth]{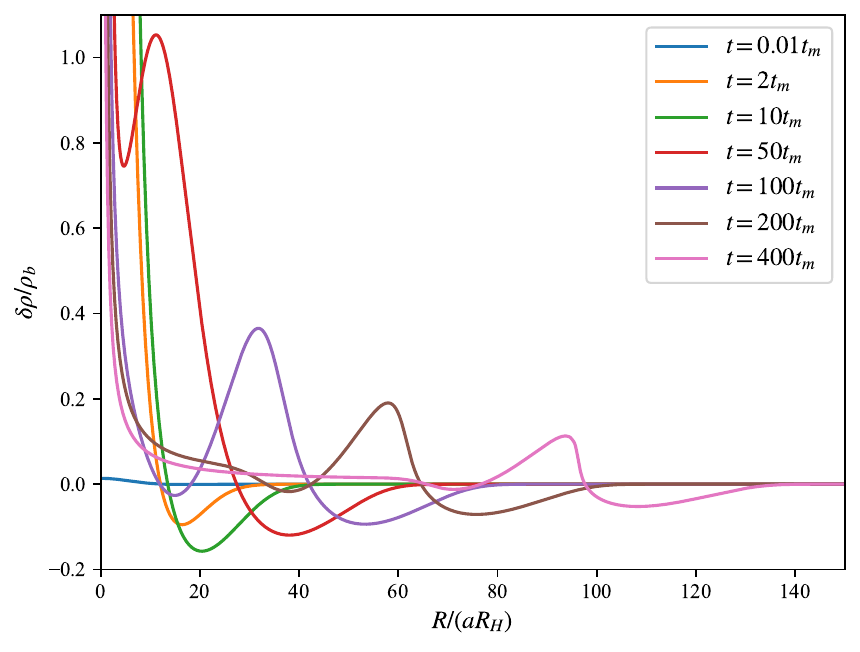}
  \qquad
  \includegraphics[width=0.45\linewidth]{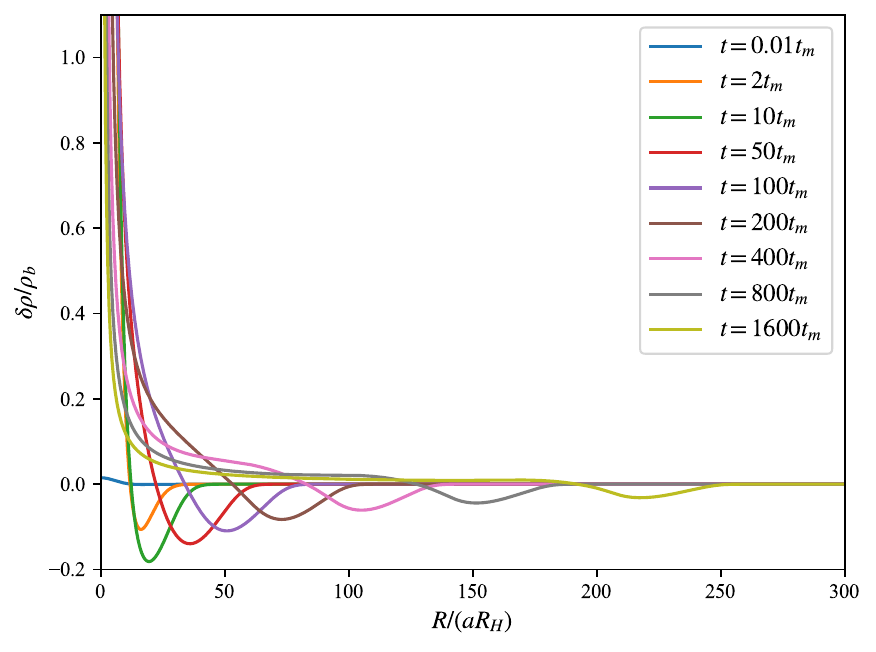}
  \caption{Time evolution of the energy density contrast for near-critical ($\delta_m = 0.5$, left panel) and far-from-critical ($\delta_m = 0.55$, right panel) perturbations.}
  \label{fig:after}
\end{figure*}

\begin{figure}[htbp]
  \centering
  \includegraphics[width=0.9\linewidth]{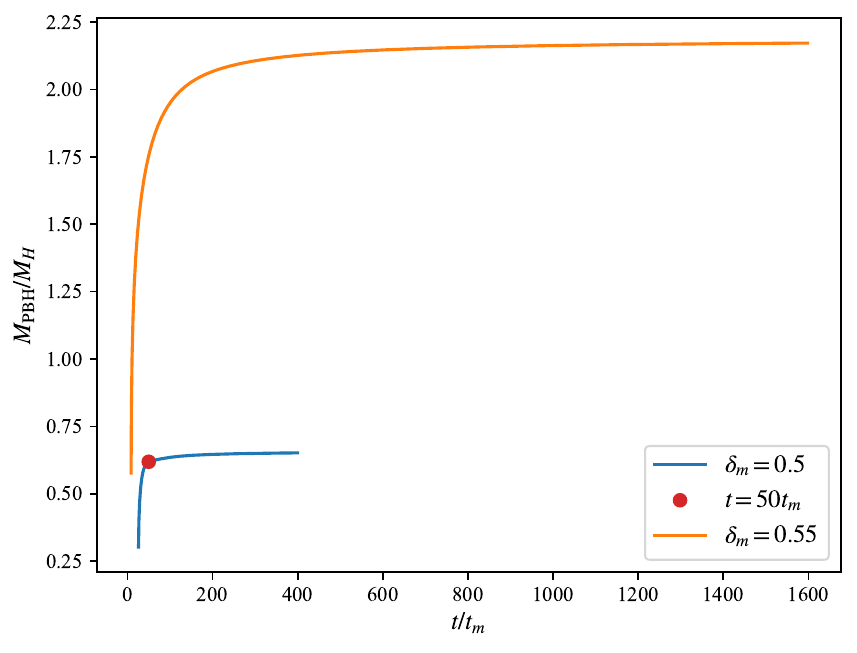}
  \caption{Time evolution of the PBH mass for near-critical ($\delta_m = 0.5$, blue line) and far-from-critical ($\delta_m = 0.55$, orange line) perturbations. The red dot marks the time ($t = 50t_m$) at which the overdense shell forms.}
  \label{fig:mass}
\end{figure}

Fig.~\ref{fig:after} displays the long-term evolution of the energy density contrast $\delta\rho \equiv \rho-\rho_b$ for the two cases. The PBH forms at $t \approx 26 t_m$ and $t \approx 8.6 t_m$, respectively. In the near-critical case, after PBH formation, material near the PBH but outside the apparent horizon bounces to form a thin overdense shell, reminiscent of the behavior observed in subcritical perturbations. This overdense shell is separated from the black hole by a rarefaction region, which effectively prevents further accretion of radiation fluid onto the PBH. The time evolution of the PBH mass for the two cases is shown in Fig.~\ref{fig:mass}. It can be seen that once the overdense shell forms (at approximately $t = 50t_m$), the PBH mass increases very slowly, reaching a final mass of roughly $2.16$ times its initial value. The energy density in the rarefaction region gradually falls below the background value, and then both the overdense shell and the initial underdensity propagate outwards, forming a compression wave. As the compression wave moves outwards, the rarefaction region dissipates, and the energy density between the PBH and the wave eventually returns to the FLRW background value.

In contrast, in the far-from-critical case, no bounce occurs, and no overdense shell is expelled. This is attributable to the formation of a larger black hole, which results in a smaller pressure gradient at the apparent horizon, preventing pressure forces from overcoming gravitational attraction. Instead, the initial underdensity propagates outwards after PBH formation and is connected to the black hole by a plateau. The absence of a rarefaction region allows for more efficient accretion, and the final PBH mass can reach up to $3.77$ times the initial mass.

Although the outward-propagating configurations differ between the two cases---with or without an overdense shell---in both instances, the energy density contrast satisfies $\delta\rho/\rho_b \ll 1$ at late times. Consequently, once these waves move sufficiently far from the PBH, they evolve into sound waves, which can be treated as linear perturbations on a flat FLRW background. The propagation of these sound waves is examined in the next subsection.

\begin{figure*}[htbp]
  \centering
  \includegraphics[width=0.45\linewidth]{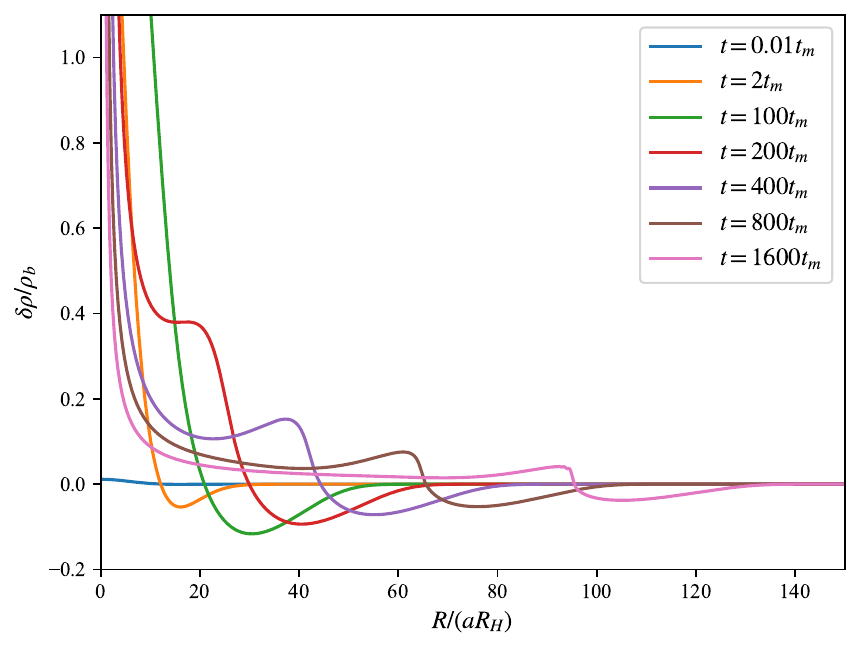}
  \qquad
  \includegraphics[width=0.45\linewidth]{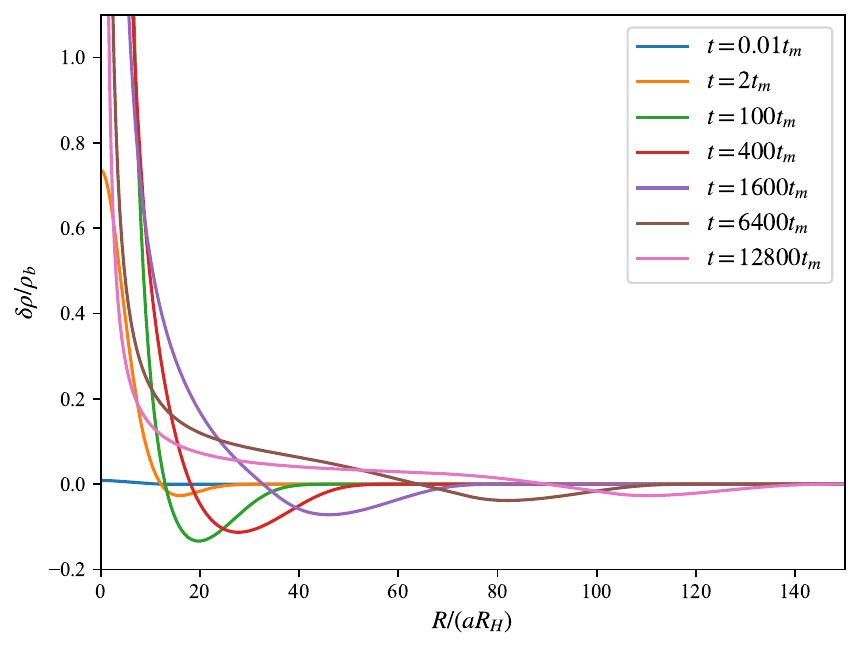}
  \caption{Time evolution of the energy density contrast for $\omega = 1/5$ (left panel) and $\omega = 1/10$ (right panel). In both cases, the perturbation amplitudes are chosen such that $\delta_m-\delta_c \approx 0.00226$.}
  \label{fig:omega}
\end{figure*}

While the radiation-dominated era provides a robust framework for PBH formation, scenarios involving phase transitions or thermal history effects~\cite{2006.16182}---such as the QCD crossover~\cite{astro-ph/9807257,astro-ph/9605152,astro-ph/9808142,1801.06138,2209.06196} or beyond-Standard-Model physics~\cite{2211.15674,2311.17760}---can significantly soften the equation of state ($\omega < 1/3$). In such cases, the PBH formation thresholds decrease substantially~\cite{1201.2379,1609.01205,2007.05564} due to reduced pressure gradients, and the generation of compression waves is expected to be more challenging. Our simulation results for $\omega = 1/5$ and $\omega = 1/10$ are presented in Fig.~\ref{fig:omega}. In both cases, the perturbation amplitudes are chosen such that $\delta_m-\delta_c \approx 0.00226$, consistent with the near-critical case for $\omega = 1/3$. The PBH forms at $t \approx 46 t_m$ and $t \approx 66.7 t_m$, respectively. It is evident that the compression wave is weaker for $\omega = 1/5$ and disappears for $\omega = 1/10$, compared to the radiation-dominated case.  This observation is consistent with the expectation that a softer equation of state---with its reduced pressure gradient---hinders pressure forces from overcoming gravity and forming compression waves.

\subsection{Propagation of sound waves}

In this subsection, we study the propagation of a spherical sound wave packet in a flat FLRW universe. The flat FLRW metric is given by
\begin{equation}
  ds^2 = -dt^2 + a^2(t)(dr^2 + r^2d\Omega^2).
\end{equation}
Here, we repurpose $r$ to denote the comoving radius, i.e., $R/a$ in the Misner-Sharp formalism. In the comoving frame, the four-velocity of the background fluid is $u_{\mathrm{bg}}^\mu = (1,0,0,0)$; however, the sound wave possesses a relative comoving velocity $u^i = \mathrm{d}x^i/\mathrm{d}t =  (u,0,0)$ (with $u^0$ determined by the normalization $u^\mu u_\mu = -1$). The energy density contrast of the sound wave is $\delta \equiv \delta\rho/\rho_b$, and both $\delta$ and $u$ are treated as first-order perturbations on the FLRW solution.

\begin{figure*}[htbp]
  \centering
  \includegraphics[width=0.45\linewidth]{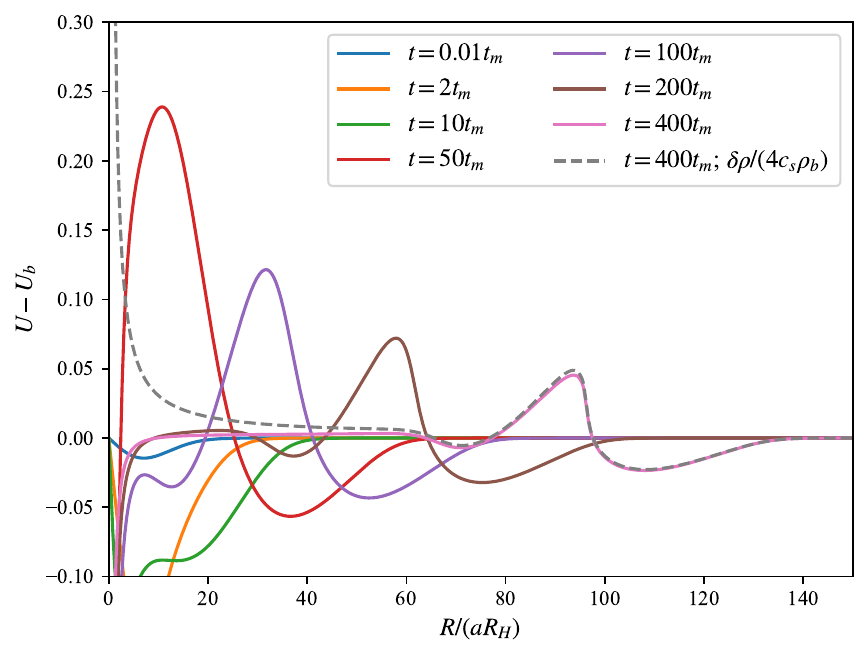}
  \qquad
  \includegraphics[width=0.45\linewidth]{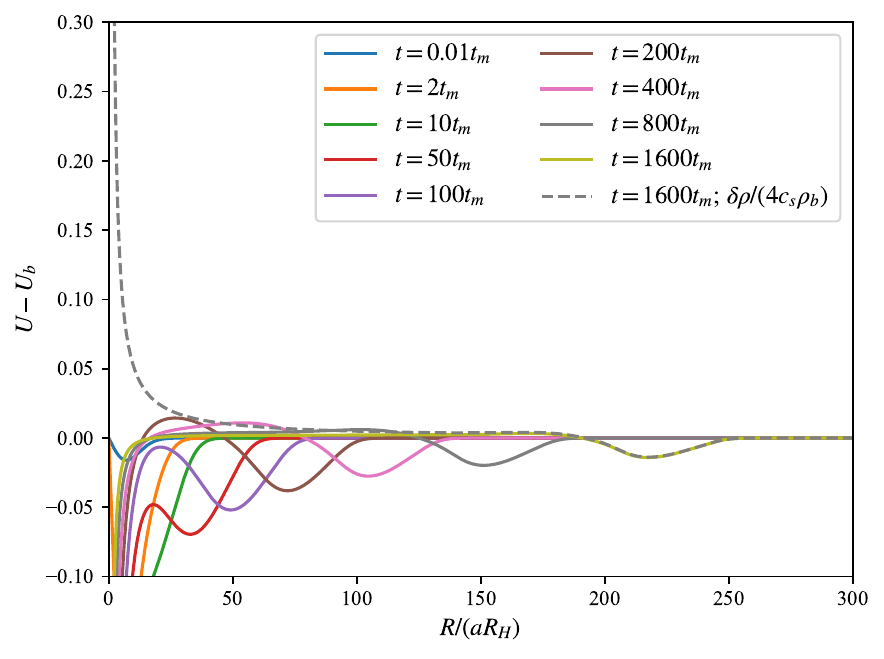}
  \caption{Time evolution of the peculiar velocity for near-critical ($\delta_m = 0.5$, left panel) and far-from-critical ($\delta_m = 0.55$, right panel) perturbations. The grey dashed line indicates the energy density contrast divided by $4c_s$ at the corresponding time.}
  \label{fig:U}
\end{figure*}

From Eq.~\eqref{eq:energy-momentum}, the energy-momentum tensor for the fluid and sound wave is
\begin{subequations}
  \begin{align}
    T^{00} &= \rho_b(1+\delta)+(1+\omega)\rho_b(1+\delta) a^2u^iu_i,\\
    T^{0i} &= (1+\omega)\rho_b(1+\delta)u^i,\\
    T^{ij} &= \omega\frac{h^{ij}}{a^2}\rho_b(1+\delta)+(1+\omega)\rho_b(1+\delta) u^iu^j,
  \end{align}
\end{subequations}
where $h^{ij}$ is the spatial metric on the unit sphere. The zeroth-order conservation equation for the energy-momentum tensor yields
\begin{equation}
  \dot{\rho}_b+3(1+\omega)\frac{\dot{a}}{a}\rho_b = 0,
\end{equation}
where the dots denote derivatives with respect to $t$. Using this relation, the linearized energy-momentum conservation equations that describe the propagation of sound waves become
\begin{subequations}
  \begin{align}
    \dot{\delta}+(1+\omega)\left(u'+\frac{2u}{r}\right) &= 0,\\
    (1+\omega)\dot{u}+\omega\frac{\delta'}{a^2}+(2-3\omega^2-\omega)\frac{\dot{a}}{a}u &= 0,
  \end{align}
\end{subequations}
where the primes denote derivatives with respect to $r$. Introducing the velocity potential $\phi$ by setting $u = \phi'$, the second equation implies that
\begin{equation}
  \delta = -\left(\frac{1}{\omega}+1\right)a^2\dot{\phi}-\left(\frac{2}{\omega}-3\omega-1\right)a\dot{a}\phi.
\end{equation}
For a radiation fluid, we set $\omega = 1/3$, so that
\begin{equation}
  \delta = -4(a^2\dot{\phi}+a\dot{a}\phi) = -4a\partial_t(a\phi).
\end{equation}
Substituting this expression into the first conservation equation, we obtain
\begin{equation}
  \partial_t\left(a\partial_t(a\phi)\right)-\frac{1}{3}\left(\phi''+\frac{2\phi'}{r}\right) = 0.
\end{equation}
This equation describes the propagation of spherical sound waves, reducing to the familiar form in Minkowski spacetime when $a \equiv 1$. For a radiation-dominated universe with $a = \sqrt{t/t_0}$, the general solution of this equation is
\begin{equation}
  \phi = \frac{f(r-c_s R_H a)}{ar},
\end{equation}
where $f$ is an arbitrary function. Consequently, the energy density contrast and the radial velocity of the sound wave are given by
\begin{subequations}
  \begin{align}
    \label{eq:delta} \delta &= -4a\partial_t(a\phi) = 4c_s\frac{f'(r-c_s R_H a)}{r},\\
    u &= \phi' = \frac{f'(r-c_s R_H a)}{ar}-\frac{f(r-c_s R_H a)}{ar^2}\\
    &\approx \frac{f'(r-c_s R_H a)}{ar} = \frac{\delta}{4ac_s},
  \end{align}
\end{subequations}
where the approximation holds when the sound wave propagates far from the PBH. In physical coordinates, the peculiar velocity of the sound wave is $U_s = au$ (i.e., the velocity with respect to the conformal time in Ref.~\cite{2007.08537}), which corresponds to $U-U_b$ in the Misner-Sharp formalism. Therefore, the relation between the peculiar velocity and the energy density contrast is
\begin{equation}
  U-U_b \approx \frac{\delta\rho/\rho_b}{4c_s},
\end{equation}
which is consistent with the result in Minkowski spacetime~\cite{2106.09817}. Our numerical results, presented in Fig.~\ref{fig:U}, confirm this relation. The figure shows the time evolution of the peculiar velocity for two perturbation amplitudes, with the gray dashed line representing the energy density contrast divided by $4c_s$. It is evident that the two quantities agree well in regions far from the PBH.

\begin{figure}[htbp]
  \centering
  \includegraphics[width=0.9\linewidth]{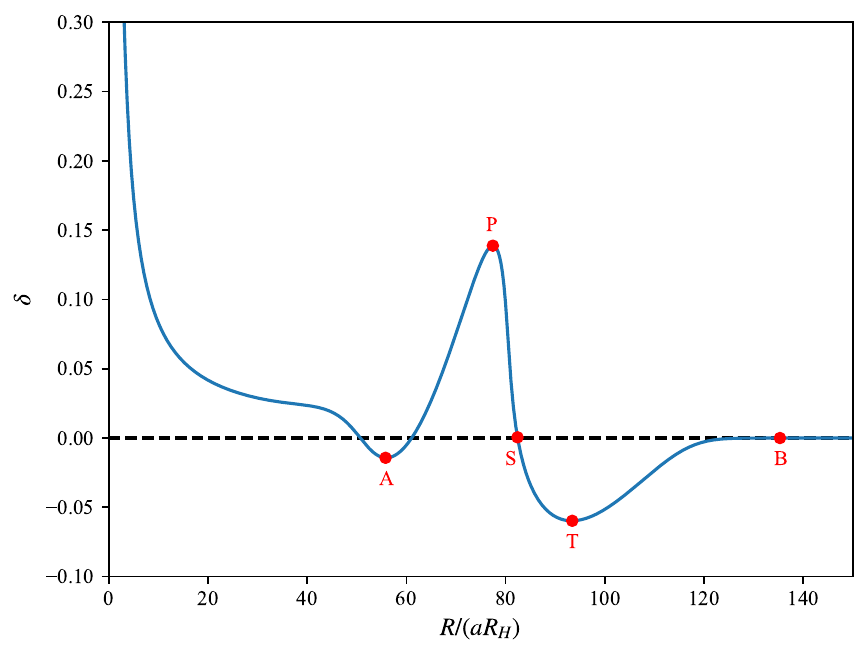}
  \caption{Schematic illustration of the sound wave profile for near-critical perturbations. Point $A$ denotes the trough of the small rarefaction region and represents the beginning of the overdense shell, while point $B$ marks a zero-crossing of the energy density contrast at the end of the underdense shell. Point $S$ is the intersection point of the two shells, and points $P$ and $T$ denote the peak and trough of the sound wave, respectively.}
  \label{fig:sketch}
\end{figure}

Eq.~\eqref{eq:delta} also reveals the propagation characteristics of sound waves in a flat FLRW universe. First, the amplitude of the sound wave (whether at a peak or trough) decays approximately as the reciprocal of the comoving radius. Second, the comoving radius of sound waves $r_S$---defined as the intersection point of the overdense and underdense shells (point $S$ in Fig.~\ref{fig:sketch})---satisfies $\Delta r_S = c_s R_H \Delta a$. In other words, in the comoving frame, the sound wave propagates at a constant speed:
\begin{equation}
  a\frac{dr_S}{dt} = c_s.
\end{equation}

Third, the sound wave as a whole propagates at the speed of sound in the comoving frame while preserving its shape. For any specific point on the wave (e.g., where the overdense and underdense shells intersect, or at the peak or trough), its comoving radius $r$ satisfies $\Delta r \approx c_s R_H \Delta a$. Consequently, the shell comoving thickness, defined as the radial distance between two specific points on the wave, remains nearly constant during propagation. Similar to the results of the sound shell model for FOPTs, our companion letter~\cite{Zeng:2025law} demonstrates that the thickness of the sound shell also determines the peak frequency of the GW spectrum in the PBH formation scenario. We will focus on the thickness of the sound shells in the next subsection.

\subsection{Thickness of sound shells} \label{sec:thickness}

In the following, we focus primarily on the near-critical case ($\delta_m = 0.5$). In this scenario, the sound wave comprises two distinct shells: an overdense shell and an underdense shell, denoted by the subscripts ``$+$'' and ``$-$'', respectively.

We define the comoving thickness of the underdense shell ($d_-$) as the distance between the two zero crossings where the energy density equals the background value (i.e., the distance between point $S$ and $B$ in Fig.~\ref{fig:sketch}). In contrast, because the overdense shell is formed by a bounce, its comoving thickness ($d_+$) is defined as the distance from the trough of the small rarefaction region to the intersection point where the energy density returns to the background value (i.e., the distance between points $A$ and $S$ in Fig.~\ref{fig:sketch}).

\begin{figure*}[htbp]
  \centering
  \includegraphics[width=0.45\linewidth]{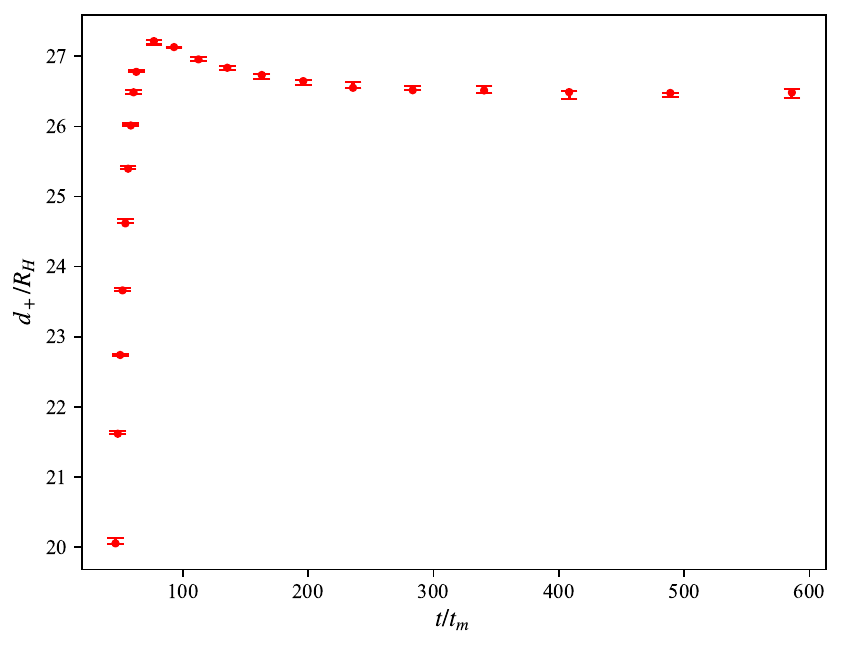}
  \qquad
  \includegraphics[width=0.45\linewidth]{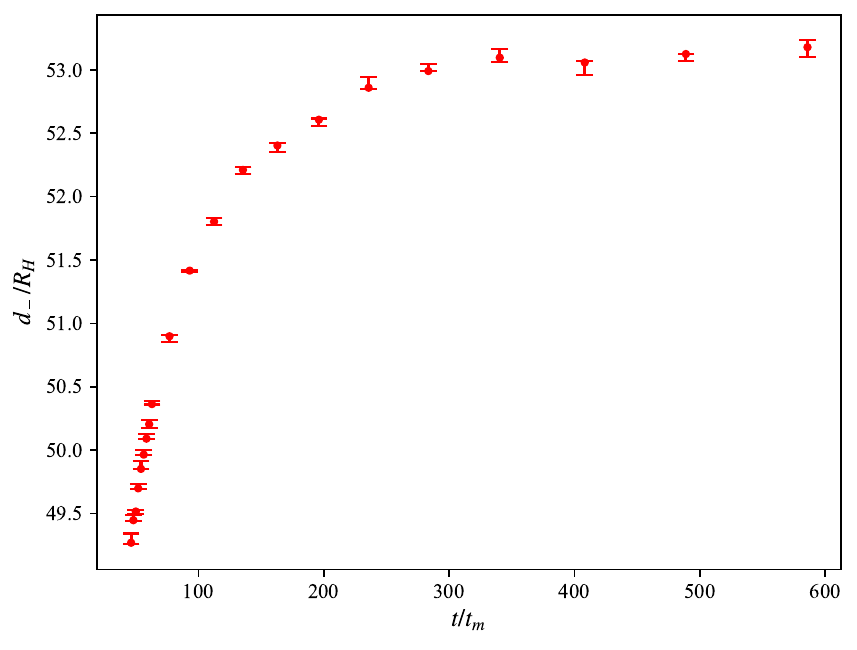}
  \caption{Time evolution of the comoving thickness of the overdense ($d_+$, left panel) and underdense ($d_-$, right panel) shells for near-critical perturbations.}
  \label{fig:thickness}
\end{figure*}

The timing of collisions between sound shells depends on the mean separation between PBHs. If collisions occur before the compression wave reaches the linear regime, the peak position of the GW spectrum may be affected. To compare these cases, Fig.~\ref{fig:thickness} illustrates the time evolution of the comoving thickness of the shells, measured from the onset of overdense shell formation (defined as the appearance of an inflection point $A$ in the energy density contrast profile). The numerical uncertainties are estimated by performing simulations with five different grid sizes ($N = 2000, 3000, 4000, 5000, 6000$). It can be seen that the comoving thickness of the overdense shell initially increases rapidly and then decreases gradually, while the comoving thickness of the underdense shell steadily increases as it propagates outwards. Nevertheless, both thicknesses vary only slightly, so it is reasonable to assume that the comoving thicknesses of the sound shells remain effectively constant.

\begin{figure}[htbp]
  \centering
  \includegraphics[width=0.9\linewidth]{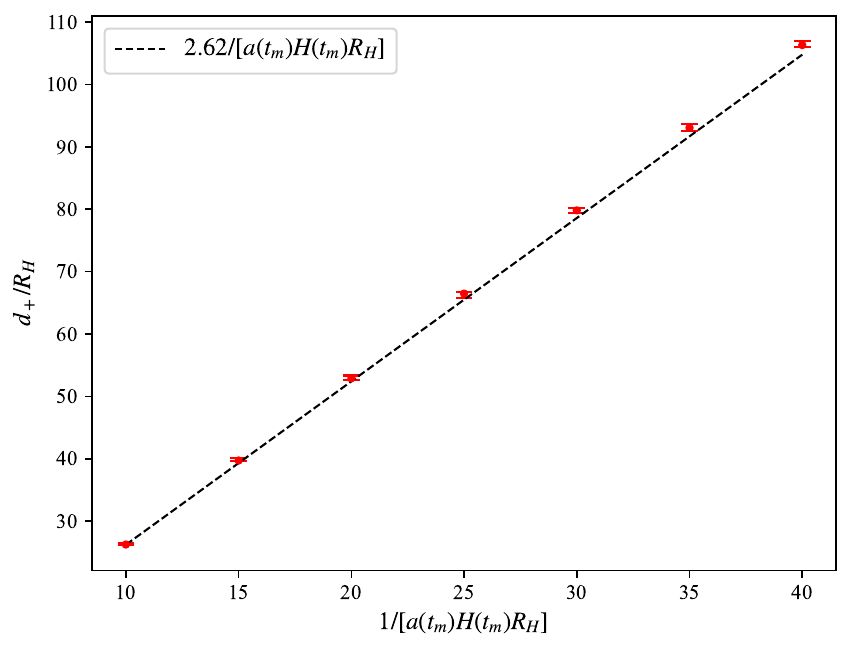}
  \caption{Comoving thickness of the overdense shell at $t = 400t_m$ as a function of the comoving Hubble radius at $t_m$ for near-critical perturbations. The black dashed line represents a linear fit.}
  \label{fig:thickness_R_H}
\end{figure}

Although the shell comoving thickness is nearly constant over time, we can expect it to be proportional to the initial perturbation length, which determines the horizon re-entry time $t_m$ and the corresponding comoving Hubble radius $1/a(t_m)H(t_m)$. On the other hand, since the mass of a PBH is of the same order as the horizon mass at the re-entry time $t_m$, relating the comoving thickness of the sound shells to the Hubble radius at $t_m$ allows one to infer the corresponding PBH mass from the GW spectrum generated by these waves. To this end, we vary the initial length scale of the perturbation to adjust its horizon re-entry time; the timescale of gravitational collapse then varies accordingly. Fig.~\ref{fig:thickness_R_H} shows the comoving thickness of the overdense shell at $t = 400t_m$ (this is a good reference time, after which the shell comving thickness remains nearly constant) as a function of the comoving Hubble radius at $t_m$, with the black dashed line representing a linear fit. The relation is given by
\begin{equation}
  d_+ \approx n/a(t_m)H(t_m),
\end{equation}
and our numerical results indicate that $n \approx 2.62$. A similar relation holds for the underdense shell, where $n \approx 5.24$. This relation is employed in our companion letter~\cite{Zeng:2025law} to estimate the present-day GW peak frequency in terms of PBH mass.

\subsection{Energy of sound shells} \label{sec:energy}

Refs.~\cite{1904.02129,1904.11482} used energy conservation to estimate the energy of the compression waves, given by:
\begin{equation} \label{eq:energy}
  E_+\simeq\left(\frac{1}{\gamma}-1\right)M_\mathrm{H}=\left(\frac{1-\gamma}{\gamma^2}\right)M_\mathrm{PBH},
\end{equation}
where $\gamma$ is the fraction of the Hubble horizon mass that collapses into a black hole (i.e., $M_\mathrm{PBH}=\gamma M_\mathrm{H}$). Although this estimate is somewhat crude, it provides a useful insight: as the perturbation amplitude approaches the critical threshold, the resulting PBH mass decreases due to the mass-scaling law, leading to a smaller $\gamma$ and, consequently, a larger compression wave energy. This trend is consistent with our numerical simulations.

To quantify the energy of the overdense shell, we use the Misner-Sharp mass, which accounts for the total energy of the spacetime, including both gravitational potential and kinetic energy. Since the Universe is radiation-dominated, the energy of the sound shells is defined as the excess energy compared to the homogeneous background. By subtracting the energy of the FLRW background, we define:
\begin{equation}
  \Delta M \equiv M-M_b = \frac{4\pi}{3}\rho_b R^3(\bar{M}-1).
\end{equation}
The energy of the overdense shell is then given by $E_+ = \int_{r_A}^{r_S} 4\pi R^2 (\rho-\rho_b) \mathrm{d}R = \Delta M(r_S)-\Delta M(r_A)$, while the energy of the underdense shell is defined as $E_- = \int_{r_S}^{r_B} 4\pi R^2 (\rho-\rho_b) \mathrm{d}R = \Delta M(r_B)-\Delta M(r_S) = -\Delta M(r_S)$, where $r_S$, $r_A$, and $r_B$ denote the comoving radii of points $S$, $A$, and $B$ in Fig.~\ref{fig:sketch}, respectively.

\begin{figure}[htbp]
  \centering
  \includegraphics[width=0.9\linewidth]{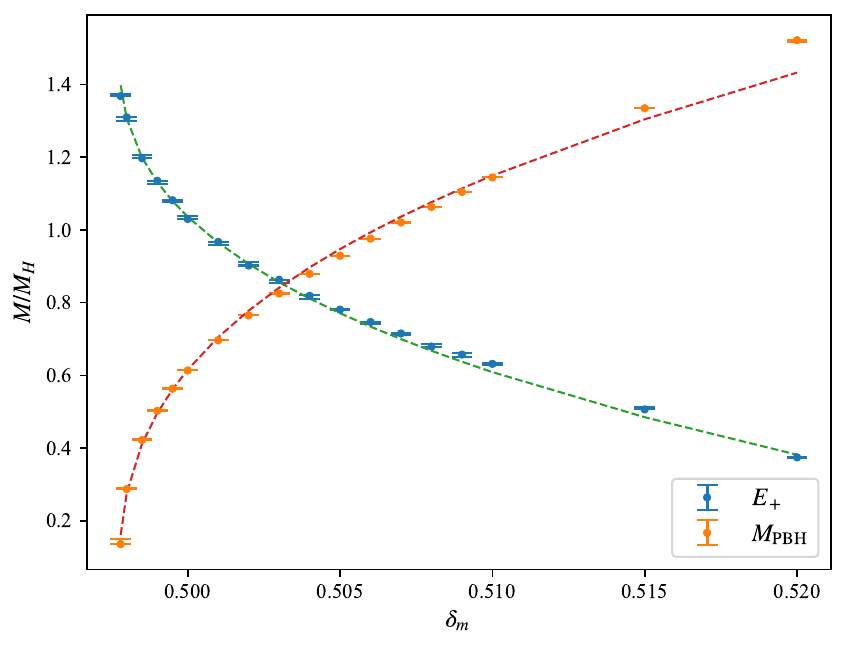}
  \caption{Energy of the overdense shell and the PBH mass at the moment of shell formation as functions of perturbation amplitude. The green and red dashed lines represent the scaling law fits for the energy of the overdense shell and the PBH mass, respectively.}
  \label{fig:M_+}
\end{figure}

Fig.~\ref{fig:M_+} shows the energy of the overdense shell and the PBH mass at the moment of shell formation (defined as the time when an inflection point $A$ appears in the energy density contrast profile) as functions of the perturbation amplitude. We choose this time rather than $t = 400t_m$ because the timescale of gravitational collapse differs for different values of $\delta_m$, whereas $t_m$ is fixed. The numerical uncertainties are also included in the figure. For $\delta_m$ near the critical threshold $\delta_c$, the PBH mass follows the scaling law~\cite{astro-ph/9901292,Hawke:2002rf,gr-qc/0412063}:
\begin{equation}
  M_\mathrm{PBH} = M_H\mathcal{K}(\delta_m-\delta_c)^\sigma,
\end{equation}
where $\mathcal{K}$ and $\sigma$ are constants. As expected, the energy of the overdense shell decreases as the perturbation amplitude deviates from the critical threshold, corresponding to a larger PBH mass. Eventually, when the perturbation amplitude exceeds a certain value, the overdense shell vanishes. Moreover, we find that the energy of the overdense shell follows a similar scaling law:
\begin{equation}
  E_+ = M_H\left[\mathcal{K}_+(\delta_m-\delta_c)^{\sigma_+}+C_+\right],
\end{equation}
where $\mathcal{K}_+$, $\sigma_+$, and $C_+$ are constants. Using the data for $\delta_m \le 0.5$, the fit yields $\sigma \approx \sigma_+ \approx 0.37$, $\mathcal{K} \approx 5.84$, $\mathcal{K}_+ \approx -4.70$, and $C_+ \approx 1.52$. In Fig.~\ref{fig:M_+}, the green and red dashed lines represent the scaling law fits for the energy of the overdense shell and the PBH mass, respectively. These fits agree well with the numerical results even for $\delta_m > 0.5$. This result suggests that the overdense shell contains at most $1.52M_H$ of energy. In Ref.\cite{1805.04087}, it was demonstrated that although critical collapse leads to a broad PBH mass spectrum, the majority of PBHs form with masses approximately $0.8M_H$. From Fig.~\ref{fig:M_+}, we observe that the energy of the overdense shell at this PBH mass is about $0.9M_H$, indicating that Eq.~\eqref{eq:energy} underestimates this energy.

Alternatively, the energy of the overdense and underdense shells can be estimated using the energy density contrast and the shell thickness:
\begin{subequations}
  \begin{align}
    E_+ &\approx 4\pi R_+^2\cdot\frac{1}{2}\delta_+\rho_b D_+ = 2\pi a^3r_+^2\delta_+\rho_b d_+,\\
    E_- &\approx 4\pi R_-^2\cdot\frac{1}{2}\delta_-\rho_b D_- = 2\pi a^3r_-^2\delta_-\rho_b d_-,
  \end{align}
\end{subequations}
where $r_+$ and $r_-$ are the comoving radii of the overdense peak and the underdense trough, respectively. These expressions align with the energy estimates in Ref.~\cite{1804.10059}, apart from an additional factor of $1/2$. Our numerical results confirm that these relations hold with a relative error of approximately $O(6\%)$. At late times, since $\delta_\pm r_\pm \sim \mathrm{const}$, $d_\pm \sim \mathrm{const}$, $r_\pm \sim \sqrt{t}$, $a \sim \sqrt{t}$, and $\rho_b \sim 1/t^2$, we find $E_\pm \sim \mathrm{const}$, which is consistent with our numerical results. However, if photon diffusion is taken into account, energy from the sound waves will be transferred to the background, leading to $\mu$-distortions in the CMB spectrum~\cite{2106.09817}.

Since we consider compensated perturbations and enforce outer boundary conditions matching the FLRW solution, the energy of the PBH must be balanced by the energy of the overdense and underdense shells, satisfying the relation
\begin{align}
  E_+ + E_- &= - \Delta M(r_A) \\
  &= -\int_{0}^{r_A} 4\pi R^2 (\rho-\rho_b) \mathrm{d}R \\
  &= -\int_{r_\mathrm{PBH}}^{r_A} 4\pi R^2 (\rho-\rho_b) \mathrm{d}R - \Delta M(r_\mathrm{PBH}) \\
  &\approx -M_\mathrm{PBH}.
\end{align}
Here $\int_{r_\mathrm{PBH}}^{r_A} 4\pi R^2 (\rho-\rho_b) \mathrm{d}R$ is the energy of the high-density region near the PBH that does not propagate outwards; its contribution to the total energy is negligible due to its small radial extent. Besides, we have $\Delta M(r_\mathrm{PBH}) = M(r_\mathrm{PBH}) - M_b(r_\mathrm{PBH}) \approx M(r_\mathrm{PBH}) = M_\mathrm{PBH}$, since the background mass within $r_\mathrm{PBH}$ is negligible compared to $M_\mathrm{PBH}$. These relations have been verified by our numerical results. Furthermore, this balance formula is generally valid across various PBH formation scenarios, as discussed in Ref.~\cite{2106.09817}. Consequently, a compensating underdense structure must always be present near PBHs, whereas the overdense structure arises only in near-critical cases.

\subsection{Effects of the curvature perturbation profile}

\begin{figure}
  \centering
  \includegraphics[width=0.9\linewidth]{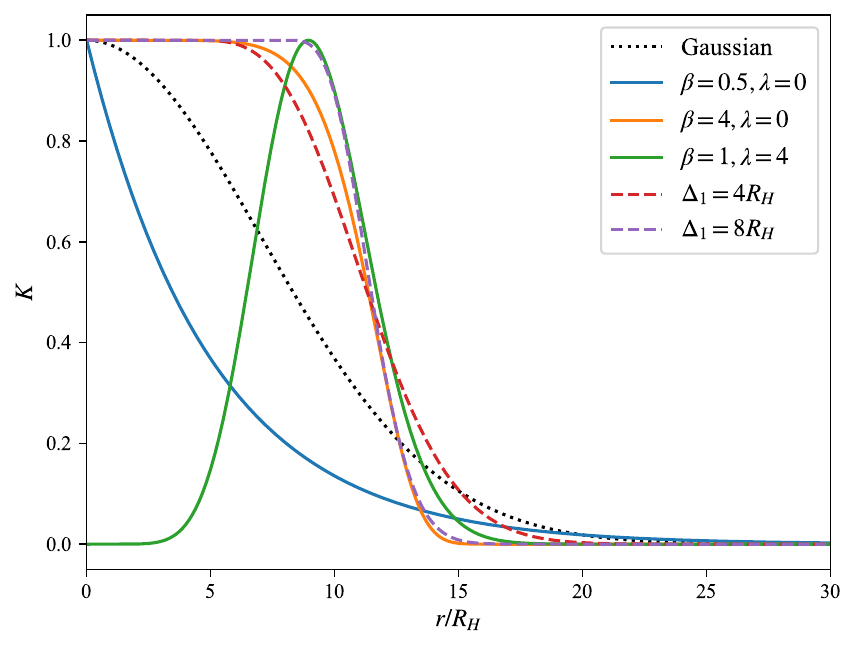}
  \caption{Curvature perturbation profiles for different cases: Gaussian profile (dotted line); generalized Gaussian profiles (solid lines) with $\beta = 0.5, \lambda = 0$ (blue), $\beta = 4, \lambda = 0$ (orange), and $\beta = 1, \lambda = 4$ (green); and smoothed top-hat profiles (dashed lines) with $\Delta_1 = 4R_H$ (red) and $\Delta_1 = 8R_H$ (purple). All profiles are normalized such that $\mathrm{max}(K) = 1$.}
  \label{fig:curvature_profile}
\end{figure}

It is well-known that the formation threshold~\cite{1310.3007,1809.02127,1907.13311,2007.05564}, mass~\cite{2103.03867}, and abundance~\cite{1805.04087} of PBHs depend on the shape of the curvature perturbations. Therefore, it is of interest to study how the formation of compression waves depends on the perturbation profile. To this end, we follow Refs.~\cite{1809.02127,gr-qc/0605122} to consider two more general profiles. First, we examine a family of generalized Gaussian profiles:
\begin{equation}
  K(r)=\mathcal{A}\left(\frac{r}{\Delta}\right)^{2\lambda}\exp\left[-\frac{1}{2}\left(\frac{r}{\Delta}\right)^{2\beta}\right],
\end{equation}
where $\Delta$ is related to the perturbation scale $r_i$ by
\begin{equation}
  \Delta = \left(\frac{2(\lambda+1)}{\beta}\right)^{-1/2\beta}r_i.
\end{equation}
Here, $\beta$ and $\lambda$ are parameters that control the shape of the perturbation: $\beta$ determines the steepness of the profile, while $\lambda$ controls the location of the peak. Eq.~\eqref{eq:perturbation} is recovered when $\beta = 1$ and $\lambda = 0$. We consider three cases: a steeper profile with $\beta = 0.5, \lambda = 0$; a flatter profile with $\beta = 4, \lambda = 0$; and an off-centered profile with $\beta = 1, \lambda = 4$. The curvature perturbation profiles for these three cases are shown as solid lines in Fig.~\ref{fig:curvature_profile}. For comparison, the Gaussian profile is also shown as a dotted line. The corresponding formation thresholds are $\delta_c \approx 0.46695 \pm 1\times10^{-5}$ (steeper), $\delta_c \approx 0.56837 \pm 1\times10^{-5}$ (flatter), and $\delta_c \approx 0.59013 \pm 1\times10^{-5}$ (off-centered).

\begin{figure*}[htbp]
  \centering
  \includegraphics[width=0.45\linewidth]{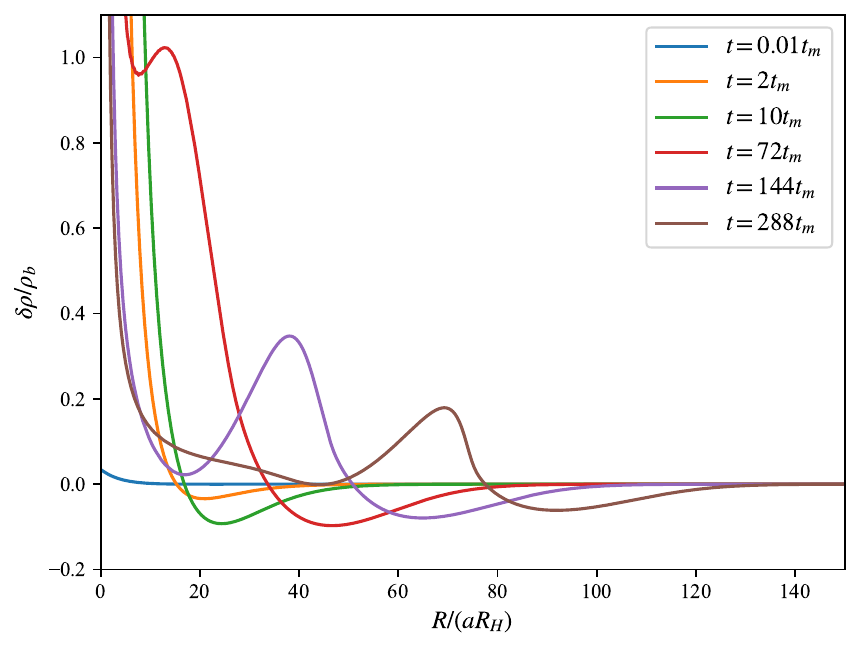}
  \qquad
  \includegraphics[width=0.45\linewidth]{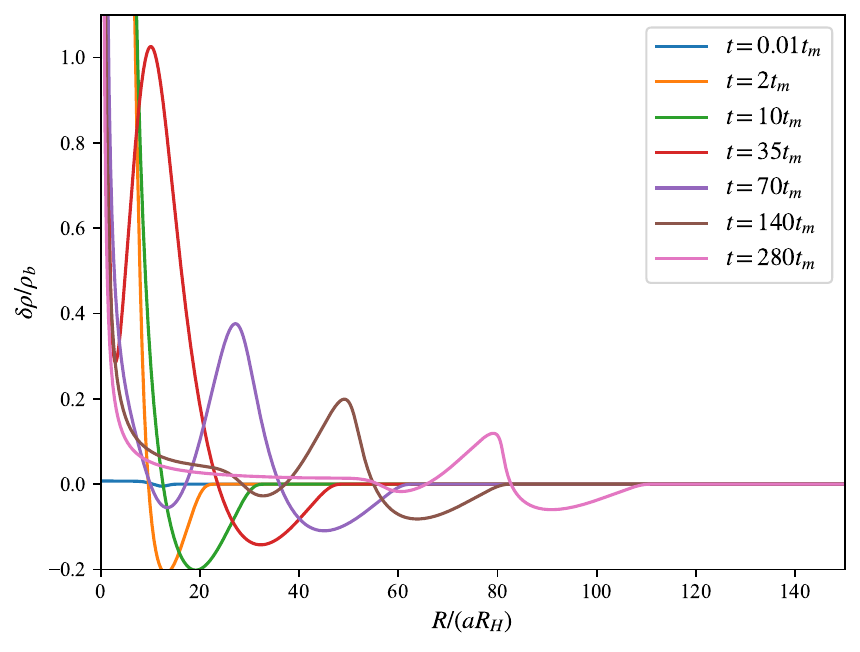}
  \\
  \includegraphics[width=0.45\linewidth]{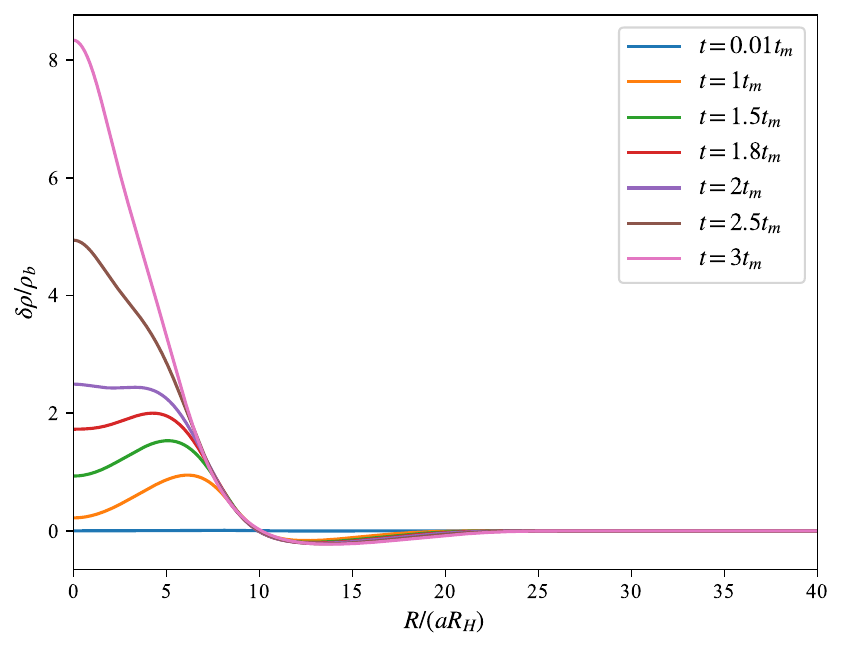}
  \qquad
  \includegraphics[width=0.45\linewidth]{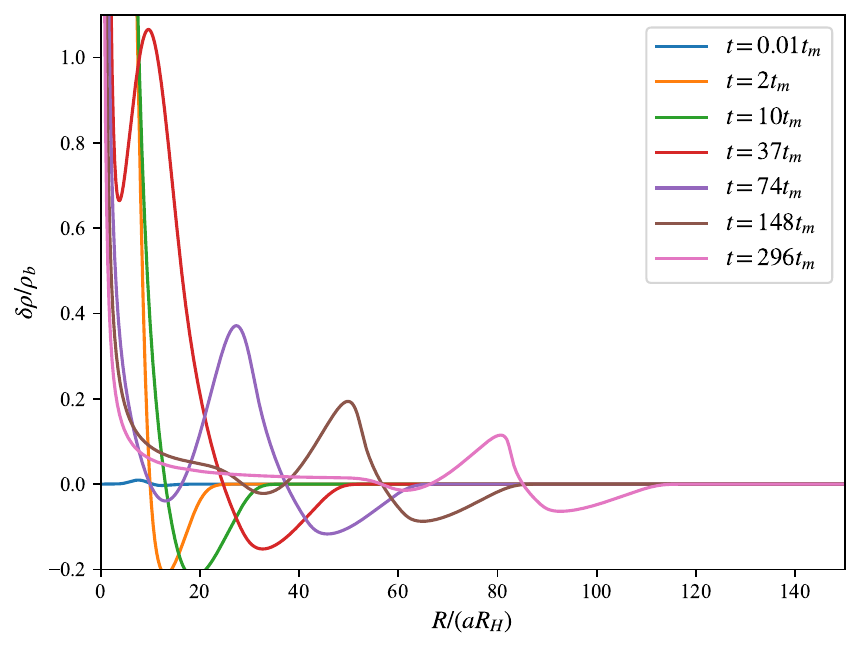}
  \caption{Time evolution of the energy density contrast for generalized Gaussian profiles: steeper profile ($\beta = 0.5, \lambda = 0$, top-left panel), flatter profile ($\beta = 4, \lambda = 0$, top-right panel), and off-centered profile ($\beta = 1, \lambda = 4$, bottom panels). For the off-centered profile, the early evolution is shown in the bottom-left panel, while the late evolution is shown in the bottom-right panel. In all cases, the perturbation amplitudes are chosen such that $\delta_m-\delta_c \approx 0.00226$.}
  \label{fig:Gaussian-like}
\end{figure*}

In all cases, the perturbation amplitudes are chosen to satisfy $\delta_m-\delta_c \approx 0.00226$ as well. Fig.~\ref{fig:Gaussian-like} shows the time evolution of the energy density contrast for these different profiles. The PBH forms at $t \approx 31.7 t_m$, $t \approx 18.7 t_m$, and $t \approx 19.8 t_m$, respectively. The flatter profile leads to a larger pressure gradient and, therefore, a higher formation threshold than the steeper profile. As can be seen in the top panel of the figure, the flatter profile produces a stronger bounce (as indicated by the red lines), reflecting an enhanced pressure gradient, while the steeper profile results in a wider compression wave. For smaller or larger $\beta$, our code is unable to follow the dynamical evolution for extended periods, and we leave the detailed study of the dependence of the compression wave shape on the perturbation profile for future work. For the off-centered profile, simulations indicate that during the early evolution, matter redistributes to fill the central depression, converging toward a centrally peaked profile (see the bottom-left panel of the figure). Consequently, the resulting compression wave is similar to that of a centrally peaked profile (as shown in the bottom-right panel of the figure).

\begin{figure*}[htbp]
  \centering
  \includegraphics[width=0.45\linewidth]{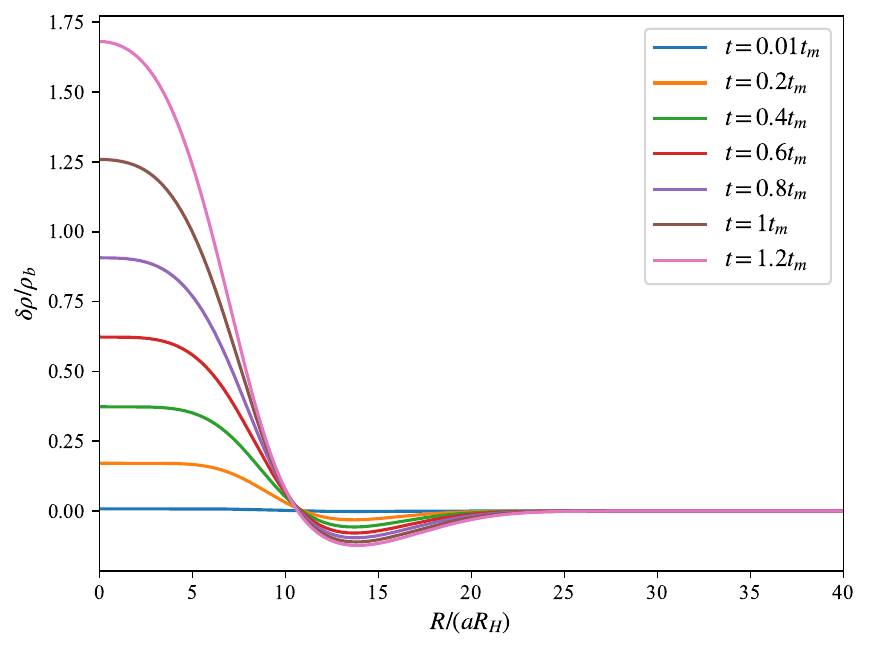}
  \qquad
  \includegraphics[width=0.45\linewidth]{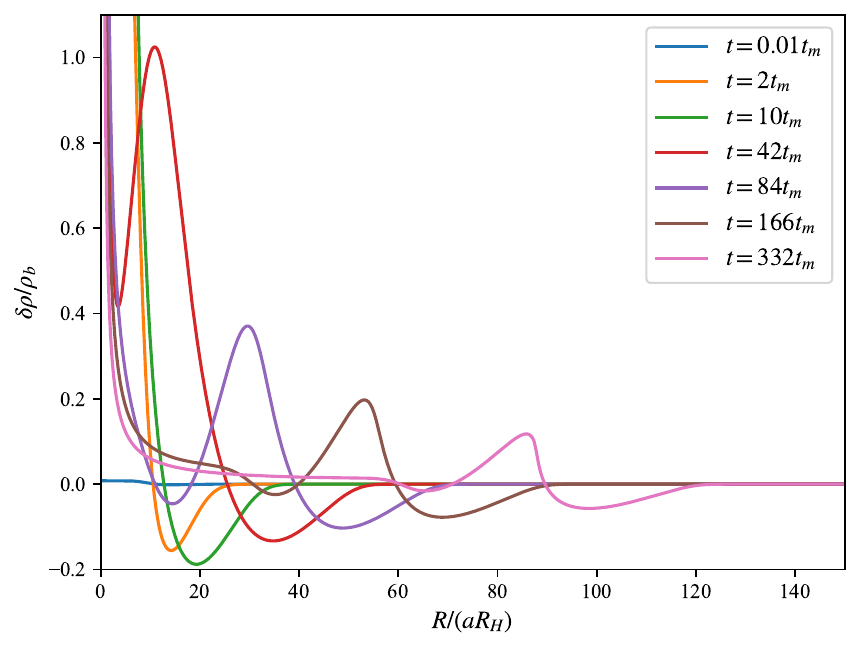}
  \\
  \includegraphics[width=0.45\linewidth]{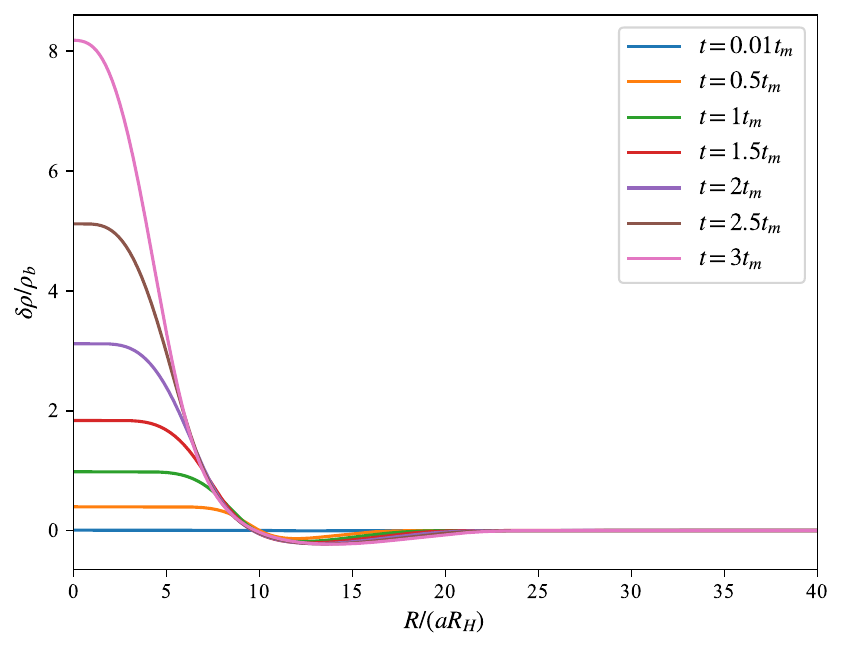}
  \qquad
  \includegraphics[width=0.45\linewidth]{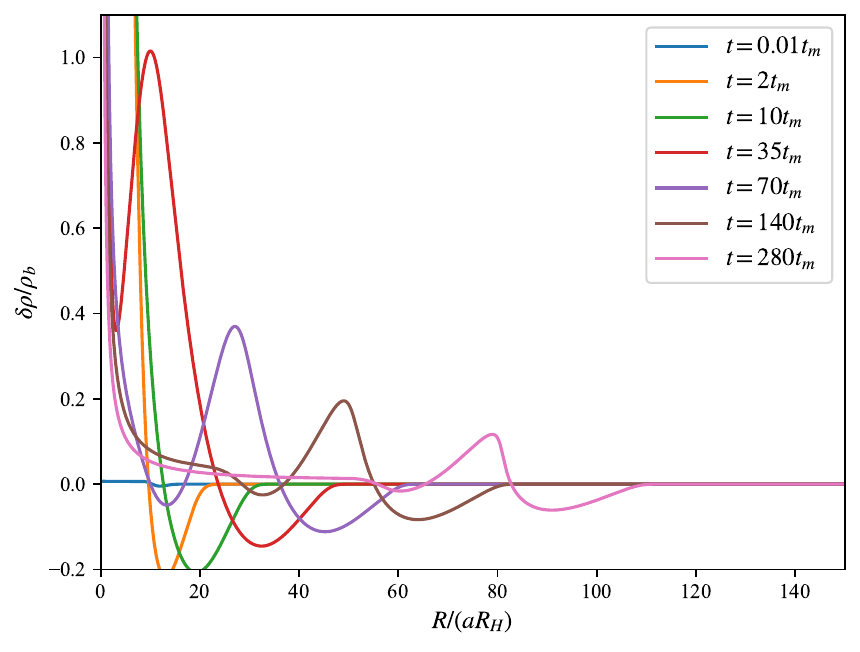}
  \caption{Time evolution of the energy density contrast for smoothed top-hat profiles with $\Delta_1 = 4R_H$ (top row) and $\Delta_1 = 8R_H$ (bottom row). The early evolution is shown in the left column, while the late evolution is shown in the right column. In all cases, the perturbation amplitudes are chosen such that $\delta_m-\delta_c \approx 0.00226$.}
  \label{fig:top-hat}
\end{figure*}

Second, we consider a smoothed top-hat profile that transitions more sharply from $\bar{K} = 1$ to $\bar{K} = 0$:
\begin{widetext}
  \begin{equation}
    K(r) = \mathcal{A}\times\left\{
        \begin{aligned}
            & 1 \quad & \text{if} \quad r \leq \Delta_1 \\
            & \left(1 + \frac{(r-\Delta_1)^2}{2\Delta_2^2}\right)\exp\left(-\frac{(r-\Delta_1)^2}{2\Delta_2^2}\right) \quad & \text{if} \quad r > \Delta_1 \\
        \end{aligned}
    \right.,
  \end{equation}
\end{widetext}
and its first derivative is
\begin{widetext}
  \begin{equation}
    K'(r) = \mathcal{A}\times\left\{
        \begin{aligned}
            & 0 \quad & \text{if} \quad r \leq \Delta_1 \\
            & -\frac{(r-\Delta_1)^3}{2\Delta_2^4} \exp\left(-\frac{(r-\Delta_1)^2}{2\Delta_2^2}\right) \quad & \text{if} \quad  r > \Delta_1 \\
        \end{aligned}
    \right..
  \end{equation}
\end{widetext}
Here, $\Delta_1$ determines the size of the top-hat region, while $\Delta_2$ controls the width of the transition region. We vary the value of $\Delta_1$ as a free parameter, while $\Delta_2$ is related to $\Delta_1$ and $r_i$ by
\begin{widetext}
  \begin{equation}
    \Delta_2 = \sqrt{- \frac{\Delta_{1}^{2}}{4} + \frac{\Delta_{1} r_i}{2} - \frac{r_i^{2}}{4} + \frac{\sqrt{\Delta_{1}^{4} - 8 \Delta_{1}^{3} r_i + 18 \Delta_{1}^{2} r_i^{2} - 16 \Delta_{1} r_i^{3} + 5 r_i^{4}}}{4}}.
  \end{equation}
\end{widetext}
We consider two cases: $\Delta_1 = 4R_H$ and $\Delta_1 = 8R_H$. The curvature perturbation profiles for both cases are shown as dashed lines in Fig.~\ref{fig:curvature_profile}. The corresponding formation thresholds are $\delta_c \approx 0.53588 \pm 1\times10^{-5}$ and $\delta_c \approx 0.58689 \pm 1\times10^{-5}$, respectively. We choose the perturbation amplitudes to satisfy $\delta_m-\delta_c \approx 0.00226$, and Fig.~\ref{fig:top-hat} displays the time evolution of the energy density contrast for these two cases. The left column shows the early evolution, while the right column shows the late evolution. During the early evolution, the top-hat region gradually shrinks as matter flows inwards, increasing the central overdensity. The PBH forms at $t \approx 22.3 t_m$ and $t \approx 18.5 t_m$, respectively. After PBH formation, a bounce occurs, leading to the formation of a compression wave similar to that observed in the near-critical case for the Gaussian profile. Comparing the two cases reveals that $\Delta_1$ has little effect on compression wave formation, as both cases exhibit similar behavior.

Finally, we measure the thickness and energy of the sound shells, as described in subsection~\ref{sec:thickness} and~\ref{sec:energy}, for the different curvature perturbation profiles considered above. The measurement time is $t = 200t_m$, after which the shell thickness and energy remain nearly constant. The results are summarized in Table~\ref{tab:summary}. Our findings suggest that the main factor influencing the properties of the sound shell is the steepness of the perturbation profiles, while the influence of offset or top-hat regions is relatively minor. Steeper profiles tend to produce thicker and more energetic sound shells.

\begin{table*}[htbp]
  \centering
  \begin{tabular}{|c|c|c|c|c|c|}
      \hline
      Profile & $\delta_c$ & $d_+/R_H$ & $d_-/R_H$ & $E_+/M_H$ & $E_-/M_H$ \\
      \hline
      Gaussian & $\makecell{0.49774 \\ \pm 1\times10^{-5}}$ & $[26.56, 26.66]$ & $[52.66, 52.83]$ & $[1.4290, 1.4325]$ & $[-2.1681, -2.1677]$ \\
      \hline
      \makecell{Gaussian-like \\ ($\beta = 0.5, \lambda = 0$)} & $\makecell{0.46695 \\ \pm 1\times10^{-5}}$ & $[33.67, 33.68]$ & $[87.99, 88.01]$ & $[2.0006, 2.0013]$ & $[-3.05432, -3.05428]$ \\
      \hline
      \makecell{Gaussian-like \\ ($\beta = 4, \lambda = 0$)} & $\makecell{0.56837 \\ \pm 1\times10^{-5}}$ & $[22.18, 22.23]$ & $[29.08, 29.20]$ & $[1.0574, 1.0579]$ & $[-1.54913, -1.54909]$ \\
      \hline
      \makecell{Gaussian-like \\ ($\beta = 1, \lambda = 4$)} & $\makecell{0.59013 \\ \pm 1\times10^{-5}}$ & $[23.75, 23.82]$ & $[34.15, 34.47]$ & $[1.1127, 1.1135]$ & $[-1.6769, -1.6766]$ \\
      \hline
      \makecell{Smoothed top-hat \\ ($\Delta_1 = 4R_H$)} & $\makecell{0.53588 \\ \pm 1\times10^{-5}}$ & $[24.11, 24.23]$ & $[39.95, 40.74]$ & $[1.2284, 1.2306]$ & $[-1.8208, -1.8203]$ \\
      \hline
      \makecell{Smoothed top-hat \\ ($\Delta_1 = 8R_H$)} & $\makecell{0.58689 \\ \pm 1\times10^{-5}}$ & $[22.33, 22.48]$ & $[30.12, 30.93]$ & $[1.0304, 1.0525]$ & $[-1.5558, -1.4913]$ \\
      \hline
  \end{tabular}
  \caption{Summary of the collapse threshold $\delta_c$, the shell comoving thickness $d_\pm$, and the shell energy $E_\pm$ for different initial curvature perturbation profiles. The measurement time for $d_\pm$ and $E_\pm$ is $t = 200t_m$. The numerical uncertainties are also included.}
  \label{tab:summary}
\end{table*}

\section{GWs from sound-wave collisions} \label{sec:GW}

Although an individual sound shell is spherically symmetric, collisions among multiple shells generate a stochastic GW background, analogous to the acoustic GW production in FOPTs. To estimate the GW energy spectrum, we adapt the sound shell model developed for FOPTs, which treats the fluid velocity as the linear superposition of many sound shells. The principal difference in our application lies in the single-shell profile. Full details and derivations are presented in our companion letter \cite{Zeng:2025law}; here we summarize the main result and display a representative spectrum.

The dimensionless GW spectrum is normalized by the critical energy density and is defined as
\begin{equation}
  \Omega_{\mathrm{GW}}(t, k) \equiv \frac{1}{\rho_c(t)}\frac{\mathrm{d}\rho_{\mathrm{GW}}(t, k)}{\mathrm{d}\ln k},
\end{equation}
where $\rho_{\mathrm{GW}}$ is the energy density of GWs and $\rho_c$ is the critical energy density of the Universe. Using the sound shell model, the GW spectrum produced by sound shell collisions is
\begin{align}
  \Omega_{\mathrm{GW}}(y, kR_H) = & 3\Gamma^2(H_sa_sR_{*c})\left(\frac{R_H}{R_{*c}}\right)^7\frac{(kR_H)^3}{2\pi^2}\nonumber\\
  &\times\tilde{P}_{\mathrm{GW}}(kR_H)\Upsilon(y),
\end{align}
where $y \equiv a/a_s$ is a time variable with the subscript $s$ denoting the time when sound waves begin sourcing GWs, $R_{*c}$ is the mean comoving separation between sound shells, $\Gamma = (\bar{\rho}+\bar{p})/\bar{\rho}\approx 4/3$, $\Upsilon$ is a suppression factor due to Hubble expansion (for radiation domination $\Upsilon \approx 1$)~\cite{2007.08537}, and $\tilde{P}_{\mathrm{GW}}$ is a convolution of the velocity power spectrum, which can be computed from the single-shell velocity profile. Further technical definitions and implementation details are provided in~\cite{Zeng:2025law}.

\begin{figure}[htbp]
  \centering
  \includegraphics[width=0.9\linewidth]{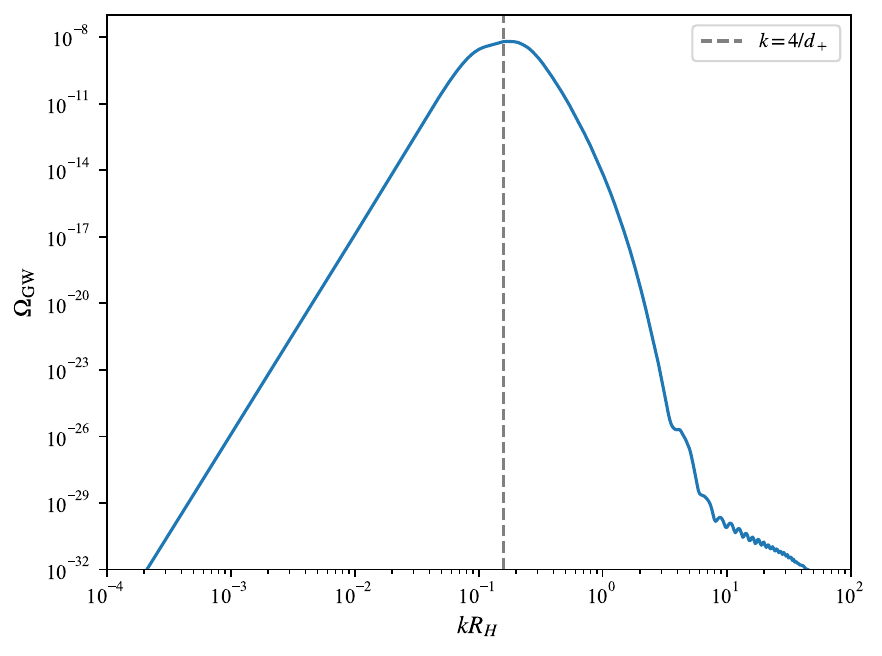}
  \caption{Dimensionless GW spectrum produced by sound shell collisions for a subcritical Gaussian perturbation with $\delta_m = 0.3$ and collision time $t_c = 50t_m$. The dashed vertical line indicates $k = 4/d_+$.}
  \label{fig:OmegaGW}
\end{figure}

As a representative example, we take a subcritical Gaussian perturbation~\eqref{eq:perturbation} with $\delta_m = 0.3$ and assume the collision time is $t_c = 50t_m$, leading to a mean comoving separation of the sound shells of approximately $R_{*c} = 134R_H$. The resulting dimensionless GW spectrum is shown in Fig.~\ref{fig:OmegaGW}, where the dashed vertical line indicates $k = 4/d_+$. The spectral peak in this example lies close to that line, but we emphasize that a rigorous analytical proof of a universal relation between the peak wavenumber and shell thickness requires further analysis of the sound shell model and is deferred to future work.

\section{Conclusions and discussion} \label{sec:conclusion}

In this work, we have numerically investigated PBH formation from super-horizon curvature perturbations and the associated generation and propagation of compression and sound waves in the early Universe. Using the Misner-Sharp formalism with an excision technique, we have explored a broad range of scenarios---from subcritical to supercritical perturbations---and examined the effects of varying equations of state and perturbation profiles.

Our results reveal that the subsequent evolution after PBH formation is highly sensitive to the perturbation amplitude. For amplitudes slightly above the critical threshold, a bounce occurs, generating a compression wave composed of both overdense and underdense shells. The rarefaction region between the PBH and the overdense shell effectively cuts off further accretion, leading to a smaller PBH mass in near-critical cases. In contrast, for significantly supercritical perturbations, only an underdense shell is observed to propagate outwards. Furthermore, we find that softer equations of state or steeper perturbation profiles inhibit the formation of compression waves, consistent with the expectation that reduced pressure gradients diminish the ability of pressure forces to overcome gravity.

As the sound shells propagate far from the PBH, they behave as linear perturbations on a flat FLRW background. The linearized conservation equations describe their propagation, demonstrating that the comoving thickness of the shells remains approximately constant---a result corroborated by our numerical simulations. Moreover, we have established an empirical relation between the comoving thickness of the shells and the comoving Hubble radius at horizon re-entry.

Furthermore, we find that the energy contained within the overdense shell, measured via the Misner-Sharp mass, decreases as the perturbation amplitude departs from the critical threshold, following a scaling law analogous to that of the PBH mass but with an opposite prefactor and an offset. Additionally, the energy of the PBH and the overdense shell must be balanced by that of the underdense shell. As a result, a compensating underdense structure is always present near PBHs, while an overdense structure forms only in near-critical cases.

Although our numerical simulations provide valuable insights into the dynamics of PBH formation and sound wave evolution, further research is needed. In this study, we have considered only generalized Gaussian and smoothed top-hat curvature perturbations. However, curvature perturbations derived from the primordial power spectrum can exhibit more complex features, such as multi-peak or oscillatory profiles, particularly in the presence of non-Gaussianities~\cite{1801.09415,1811.07857,1905.13202,1712.09998,1908.11357,2202.01028}. These variations may significantly modify sound wave profiles, leading to different GW spectra---a topic that merits further investigation. Future studies should also examine non-spherical effects~\cite{2004.01042,2403.11147,2410.03451,2410.03452} and alternative PBH formation mechanisms to assess the robustness of our conclusions across a wider range of cosmological scenarios.

Moreover, simulating the formation of multiple PBHs and the interactions between their associated sound waves would offer a critical test of our proposed GW generation mechanism and the applicability of the sound shell model. This scenario, reminiscent of the black hole lattice universe~\cite{1204.2411,1306.1389,1404.1435,1204.3568,1312.0494,1402.3201,1705.01892,1801.01083,1811.00762,2012.14049}, is left for future work.

In summary, our study provides a comprehensive analysis of PBH formation and the subsequent evolution of sound waves, shedding new light on the interplay between nonlinear gravitational collapse, fluid dynamics, and early Universe cosmology. These results enhance our understanding of PBH formation and pave the way for future research into their observable cosmological consequences.

\begin{acknowledgments}
We thank an anonymous referee for greatly improving the robustness of this paper. This work is supported by the National Key Research and Development Program of China Grants No. 2021YFC2203004, No. 2021YFA0718304, and No. 2020YFC2201501, the National Natural Science Foundation of China Grants No. 12422502, No. 12547110, No.12588101, No. 12235019, and No. 12447101, and the Science Research Grants from the China Manned Space Project with No. CMS-CSST-2025-A01 (supported by China Manned Space Program through its Space Application System). HD was supported by Yuri Levin's Simons Investigator Grant PG012519.
\end{acknowledgments}

\appendix

\section{Convergence test} \label{app:convergence}

In this appendix, we present a convergence test for our numerical code. To quantify the numerical error, we use the $L_2$-norm of the Hamiltonian constraint equation~\eqref{eq:Hamiltonian}, which should vanish if the Einstein and conservation equations are solved correctly. Specifically, we define the Hamiltonian constraint violation and its $L_2$-norm as
\begin{subequations}
  \begin{align}
    \mathcal{H}&\equiv\frac{D_rM}{4\pi\Gamma\rho R^2}-1,\\
    {||\mathcal{H}||}_2&\equiv\frac{1}{N}\sqrt{\sum_{i=1}^N{|\mathcal{H}_i|}^2}.
  \end{align}
\end{subequations}
In terms of our redefined variables, the Hamiltonian constraint violation is expressed as
\begin{equation}
  \mathcal{H} = \frac{\bar{M}+\frac{\bar{r}\bar{R}}{3(\bar{r}\bar{R})'}\bar{M}'}{\bar{\rho}}-1.
\end{equation}
To eliminate errors arising from the use of the long-wavelength approximation for the initial conditions, we subtract the initial Hamiltonian constraint violation at each time step.

\begin{figure*}[htbp]
  \centering
  \includegraphics[width=0.45\linewidth]{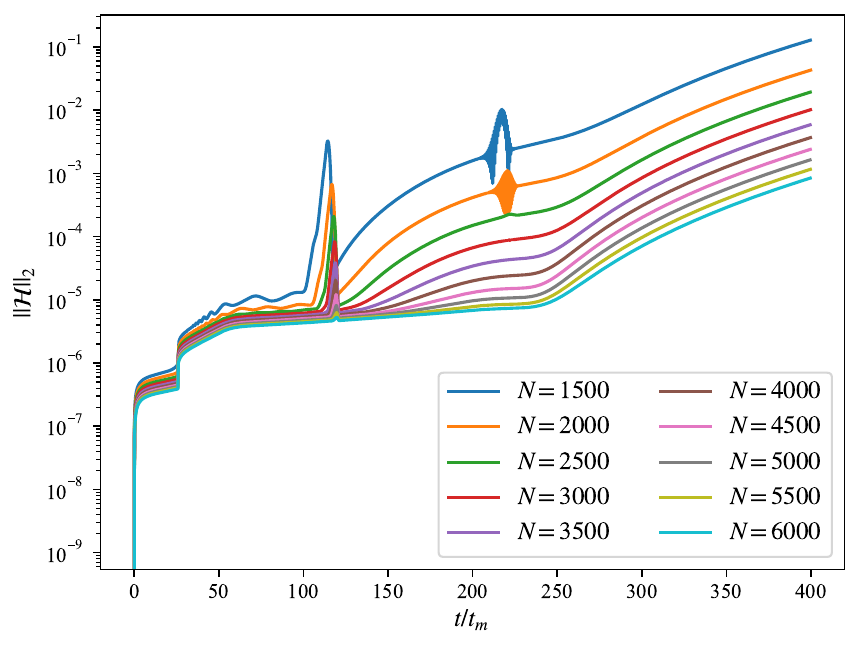}
  \qquad
  \includegraphics[width=0.45\linewidth]{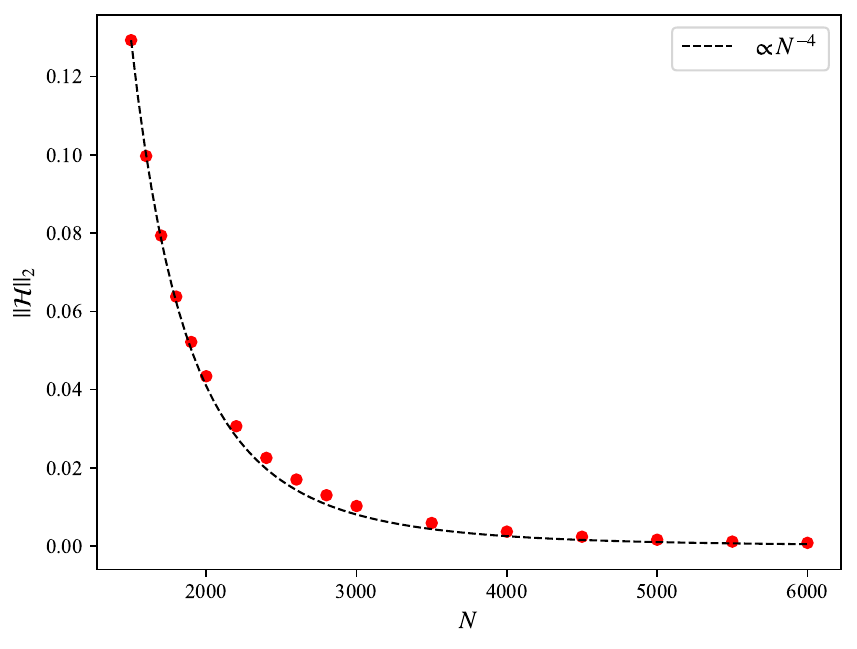}
  \caption{Convergence test of the numerical code. \textit{Left panel}: Time evolution of the $L_2$-norm of the Hamiltonian constraint violation for various grid resolutions. \textit{Right panel}: $L_2$-norm of the Hamiltonian constraint violation at the end of the simulation as a function of the grid resolution. The black dashed line indicates the expected $N^{-4}$ scaling.}
  \label{fig:convergence}
\end{figure*}

We use the near-critical perturbation ($\delta_m = 0.5$) as a test case. The code is run at various grid resolutions, with $N = 1500,2000,2500,3000,3500,4000,4500,5000,5500,6000$, and the logarithmic grid constant is set to $C = 1+9/N$ for all cases. All other parameters are identical to those used in the main text. The left panel of Fig.~\ref{fig:convergence} shows the time evolution of the $L_2$-norm of the Hamiltonian constraint violation. It can be seen that, for sufficiently high grid resolution, the violation remains at approximately $10^{-3}$ throughout the simulation (extending much later than in previous studies). The right panel of Fig.~\ref{fig:convergence} plots the $L_2$-norm at the end of the simulation as a function of the grid resolution. The black dashed line, representing a $N^{-4}$ scaling, fits the data well, confirming the fourth-order convergence of the central difference method employed.

\bibliography{biblio.bib}

\end{document}